\documentclass[journal]{IEEEtran}
\usepackage{times,amsmath,amssymb,amsfonts,amsthm, graphicx,epsfig,cite,psfrag}
\usepackage{algorithmicx,algorithm}
\usepackage{color}
\usepackage{algpseudocode}
\usepackage{enumitem}
\def\alphabf{\boldsymbol \alpha}

\def\gammabf{\boldsymbol \gamma }

\def\deltabf{\boldsymbol \delta }

\def\lambdabf{\boldsymbol \lambda }
\def\mubf{\boldsymbol \mu }
\def\nubf{\boldsymbol \nu }

\def\abf{{\bf a}}

\def\dbf{{\bf d}}

\def\fbf{{\bf f}}

\def\hbf{{\bf h}}

\def\qbf{{\bf q}}

\def\ubf{{\bf u}}

\def\wbf{{\bf w}}
\def\xbf{{\bf x}}

\def\zbf{{\bf z}}

\def\xbf{{\bf x}}

\def\Abf{{\bf A}}

\def\Cbf{{\bf C}}
\def\Dbf{{\bf D}}
\def\Ebf{{\bf E}}

\def\Gbf{{\bf G}}
\def\Hbf{{\bf H}}
\def\Ibf{{\bf I}}

\def\Rbf{{\bf R}}

\def\Wbf{{\bf W}}

\def\Zbf{{\bf Z}}

\def\Cc{{\cal C}}
\def\Dc{{\cal D}}

\def\Fc{{\cal F}}

\def\Jc{{\cal J}}
\def\Kc{{\cal K}}
\def\Lc{{\cal L}}

\def\Nc{{\cal N}}

\def\Pc{{\cal P}}

\def\ie{{\it i.e.,\ \/}}
\def\nn{\nonumber}

\def\tr{\textrm{tr}}
\def\st{\textrm{s.t.}}

\def\SCA{\textrm{SCA}}
\def\ADMM{\textrm{ADMM}}
\def\SINR{\textrm{SINR}}

\def\wrt{\textrm{w.r.t.}}
\theoremstyle{definition}

\newtheorem{theorem}{Theorem}
\newtheorem{lemma}{Lemma}

\newtheorem{remark}{Remark}
\newtheorem{corollary}{Corollary}

\newenvironment{mylist}%
{\begin{list}{}%
    {%
      \setlength{\itemindent}{-5pt}%
      \setlength{\leftmargin}{12pt}%
      \setlength{\parsep}{\parskip}
      \setlength{\labelsep}{5pt}
      \setlength{\itemsep}{2pt}}}%
  {\end{list}}

\begin{document}

\title{Computation-and-Communication Efficient  Coordinated  Multicast Beamforming in   Massive MIMO Networks }

\author{\IEEEauthorblockN {Shiqi Yin and Min Dong, \textit{Fellow, IEEE}}\thanks{The authors are with the Department of Electrical, Computer and Software Engineering, Ontario Tech University, Ontario, Canada (e-mail: shiqi.yin@ontariotechu.net, min.dong@ontariotechu.ca).}}
\maketitle

\begin{abstract}
The main challenges in designing downlink coordinated  multicast beamforming
 in massive  multiple-input multiple output (MIMO) cellular networks are the complex computational solutions and significant fronthaul overhead for centralized coordination. This paper proposes a coordinated multicast beamforming solution  that is both computation and communication efficient. For joint BS coordination with individual base station transmit power budgets, we first obtain the optimal structure of coordinated multicast beamforming.  It reveals that the  beamformer at each BS is naturally distributed and only depends on the local channel state information (CSI) at its serving BS. Moreover, the optimal beamformer  is a weighted minimum mean square error (MMSE) beamformer with  a low-dimensional structure of unknown weights  to be optimized, independent of the number of BS antennas. Utilizing the optimal structural properties, we propose fast algorithms to determine the unknown parameters
for the optimal beamformer. The main iterative algorithm decomposes the problem into small subproblems, yielding only closed/semi-closed form updates.  Furthermore, we propose a semi-distributed computing approach  for the proposed algorithm that allows each BS to compute its beamformer based on the local CSI without the need for global CSI sharing, resulting in  the fronthaul overhead  independent of the number of BS antennas.
We further extend our results to the design under the imperfect CSI and other coordination scenarios. Simulation results demonstrate that our proposed methods can achieve near-optimal performance with significantly lower computational time for massive MIMO systems than the conventional approaches.
\end{abstract}

%\begin{IEEEkeywords} Coordinated multicast beamforming, optimal beamforming %structure, distributed beamforming, distributed computing, computational %complexity
%\end{IEEEkeywords}

\section{Introduction} \label{sec:Introduction}
Data distribution and sharing have become increasingly common in the rapidly growing  wireless applications and  emerging computing paradigms. Many wireless services involve the distribution of shared content to mobile users.
Additionally, distributed machine learning (ML) through collaboration among devices over wireless networks has emerged as a promising approach for  intelligent network management  to support applications such as edge computing, the Internet of Things (IoT), and the next generation of wireless networks \cite{Zhuetal:TCM20}. Promising techniques, such as federated learning and edge learning \cite{Limetal:COMST20}, require frequent distribution of the global model update to devices, which is expected to generate significant network data traffic, particularly since ML models are often large. For this type of data, wireless multicast can play a critical role in efficiently reducing unnecessary data traffic and overhead over cellular networks \cite{Araniti&etal:Network17, Zhang&etal:WiOpt2023,Zhang&etal:ICASSP2024,Kalarde&etal:2024}.

 For data transmission at the base stations (BSs), multicast beamforming is an efficient multi-antenna  technique that enables simultaneous transmission of common data to multiple users or devices without causing interference among them. It is  both bandwidth and power efficient, making it a valuable physical-layer solution for data distribution. To further enhance the efficiency of data multicasting in cellular networks, cooperation among BSs is crucial. However,  in practical scenarios where data cannot be shared among BSs -- such as with delay-sensitive data, limitations in fronthaul for large data transfers, or difficulties in achieving strict synchronization among BSs -- coordinated multicast beamforming  is an effective approach. It combines multicast beamforming with BS coordination to effectively manage inter-cell interference and improve data multicasting efficiency.
However, the improvement comes at the cost of increased fronthaul overhead, as joint BS coordination is a centralized process that requires global channel state information (CSI) sharing among BSs. With limited fronthaul capacity, this requirement   can become a bottleneck  in  massive multiple-input multiple-output (MIMO)\ networks. Therefore, to enable effective coordination in next-generation massive MIMO cellular networks, it is critical to develop a practical solution that not only delivers high performance but is also scalable to the network size, with low computational complexity and fronthaul overhead.

However, multicast beamforming design  is generally a  challenging and complicated problem, even in a single-cell scenario, due to the NP-hard nature of the core problem \cite{Sidiropoulosetal&Davidson:SP06,Tranetal&Hanif:SP14,KonarSidiropoulos:TSP17,Karipidisetal&Sidiropoulos:SP08,Christopoulosetal&Chatzinotas:SP14,Changetal&Luo:sp08,Christopoulosetal&Chatzinotas:SPAWC15}.  Most of the existing literature on downlink multicast beamforming has been  focused on a single-cell scenario for either  a single  group  \cite{Sidiropoulosetal&Davidson:SP06,Tranetal&Hanif:SP14,KonarSidiropoulos:TSP17} or  multiple  groups \cite{Karipidisetal&Sidiropoulos:SP08,Christopoulosetal&Chatzinotas:SP14,Changetal&Luo:sp08,Christopoulosetal&Chatzinotas:SPAWC15}.
These problems are non-convex and NP-hard; thus, computational methods for approximate solutions have been  sought. For the traditional multi-antenna systems, semi-definite relaxation (SDR) has been the popular method  to obtain a good approximate solution  \cite{Sidiropoulosetal&Davidson:SP06,Karipidisetal&Sidiropoulos:SP08,Changetal&Luo:sp08,Christopoulosetal&Chatzinotas:SP14}. While it  performs well for small problems, SDR faces high computational complexity and deteriorating performance as   the number of antennas and users increases. To address this issue, successive convex approximation (SCA)  has emerged as a more appealing approach, offering improved performance and computational efficiency compared to SDR \cite{Tranetal&Hanif:SP14,Christopoulosetal&Chatzinotas:SPAWC15}. Despite the complexity reduction, these  methods  are still not scalable   for  massive MIMO systems, where BSs are typically equipped with a large number of antennas. To tacking this issue,    a zero-forcing based processing  scheme \cite{Sadeghietal&Sanguinetti:TWC17} and fast optimization-based computational algorithms \cite{KonarSidiropoulos:TSP17,Chen&Tao:COM17} have been proposed to further reduce the computational complexity of the SCA method. Also,  low-complexity robust multicast beamforming  algorithms under  the CSI uncertainty are  proposed \cite{Mohamadietal:TSP22,Mohamadietal:TSP24}. Multicasting  transmission aided by a reconfigurable intelligent surface (RIS) has also been studied in \cite{Zhouetal:TSP21} using the majorization-minimization approach to reduce the complexity of obtaining the RIS reflection coefficients.     In contrast to these computational optimization approaches, the optimal structure of multi-group multicast beamforming in the single-cell case has
  been obtained recently in \cite{Dong&Wang:SP20}. It shows that the optimal beamformer  has an inherent low-dimensional structure, where the number of unknowns 
to be computed is independent of the number of BS antennas. Based on this structure, several first-order fast algorithms have been developed, providing high computational efficiency that is suitable for large-scale massive
MIMO systems \cite{Zhangetal&Dong:WLC22,Zhangetal&Dong:SP23,Mohammadi&Dong&SS:TSP21,EbrahimiDong:Asilomar23,Li&Liu:TCOM24}. These efficient algorithms have been employed in downlink and uplink beamforming for maximizing federated learning performance \cite{Zhang&etal:WiOpt2023,Zhang&etal:ICASSP2024,Kalarde&etal:2024,Kalardeestal:MSWIM2023}

Despite these advancements, the previous studies have been primarily focused on single-cell scenarios. Studies on multicast beamforming design in multi-cell scenarios are relatively limited and mostly pertain to   traditional multi-antenna systems \cite{Jordanetal&Gong:globecom09,Dartmann&etal:Crowncom11,Xiangetal&Tao:WCOM13}.  The works in \cite{Jordanetal&Gong:globecom09} and \cite{Dartmann&etal:Crowncom11} consider full data sharing and full cooperation among BSs under a total power budget of all BSs,  which is similar to the single-cell multi-group multicast case. In  \cite{Xiangetal&Tao:WCOM13}, coordinated multicast beamforming to manage  inter-cell interference is considered, where  a decentralized SDR-base method  is proposed to minimize the total power consumed by all BSs. However, this total power constraint is often unrealistic  in practice for individually operated BSs, and the  proposed algorithm is not scalable for massive MIMO systems. Low-complexity coordinated multicast beamforming design in massive MIMO cellular networks is investigated in \cite{Yu&Dong:ICASSP18} and \cite{Yu&Dong:SPAWC18}. These studies propose  weighted maximum ratio transmission (MRT) beamforming schemes in combination with SDR to maximize the minimum signal-to-interference-and-noise ratio (SINR) among users. While these schemes  aim to reduce the solution complexity by using a suboptimal beamforming scheme, they require fully centralized processing for coordination. The communication overhead is not addressed in these works.
Due to the complexity associated with the core problem of multicast beamforming, there are few  efficient coordinated multicast beamforming designs suitable for massive MIMO cellular networks, particularly in terms of both computational complexity and fronthaul overhead required.

Existing designs for coordinated multicast beamforming  in the literature primarily rely on computational methods or specific suboptimal beamforming schemes. As previously mentioned, in the single-cell scenario, the obtained optimal  structure of multicast beamforming has led to the development of highly efficient algorithms for massive MIMO. This raises important questions  regarding the optimal structure of coordinated multicast beamforming in multi-cell scenarios and whether it can be leveraged to improve  design efficiency. These questions  remain largely unexplored in  existing literature. Understanding the optimal beamforming  structure and its inherent relationships among the coordinating cells is crucial not only for  improving our theoretical understanding but also  for  developing scalable  solutions for  massive MIMO networks. Gaining this structural insight is important for tackling  both computational complexity and significant fronthaul overhead for sharing global CSI to enable centralized coordination among BSs. Driven by these challenges and potential opportunities,  this paper aims to study  the  optimal structure of coordinated multi-cell multicast beamforming. The goal is to develop low-complexity  multicast beamforming solutions that also have low fronthaul overhead for BS coordination in massive MIMO networks.

%%%%%%%%%%%%%%%%%%%%%%%%%%%%%%%%%%%%%%%%%%%%%%%%%%%%%%%%%%%%%%%% 
\subsection{Contribution}\label{sec:reWo}
%%%%%%%%%%%%%%%%%%%%%%%%%%%%%%%%%%%%%%%%%%%%%%%%%%%%%%%%%%%%%%%%
To study  the BS coordination for multicasting, we focus on
the  quality-of-service (QoS) beamforming design formulation, aiming   to minimize each BS transmit power while meeting the user SINR targets.
We obtain the optimal structure of coordinated multicast beamforming  and then utilize it to develop a fast and scalable semi-distributed algorithm to allow each BS to compute beamformers based on the local CSI without global CSI\ sharing, thereby achieving both computation and communication efficiency. Our contribution is summarized below. 

\begin{itemize}
\item  Our QoS problem formulation  aims to minimize  each BS transmit power margin relative to its own power budget, which is more practical than  the total  power consideration  in previous works. We
derive the optimal coordinated multicast beamforming structure  by combining the SCA   properties and the Lagrangian duality.
The optimal structure reveals two essential properties: First, the optimal coordinated multicast beamformers are naturally distributed, relying only on  the local CSI  at their respective serving BSs.
Second, the optimal multicast beamformer   is a weighted minimum mean squared error (MMSE) beamformer, with the unknown weights  to be determined based on the number of serving users,  independent of the number of BS antennas. These structural properties are  particularly valuable for developing  efficient algorithmic solutions for  massive MIMO networks.

\item We propose our  fast algorithms based on the optimal solution structure. In particular, we develop a first-order fast iterative algorithm to compute the unknown weights in each optimal beamformer based on SCA and the alternating direction method of multipliers (ADMM) construction. Our ADMM construction decomposes the joint optimization problem into small per-BS or per-user subproblems, yielding   closed-form or semi-closed-form  iterative updates.
Furthermore, we propose a semi-distributed computing approach to perform the algorithm between the BSs and the central processing unit (CPU). This approach only requires essential information to be shared with the CPU, while each BS uses local CSI to compute its beamformer. It eliminates the need for global CSI sharing, resulting in the fronthaul overhead to be independent of the number of BS antennas and increase only quadratically with the total number of users in the coordinating cells.

\item We further consider the coordination design under the imperfect CSI and extend our results, including the optimal structural properties and semi-distributed fast algorithm, to this case. Generalization to other coordination scenarios, such as BS clustering is also considered.
        
\item Simulation results show that  our proposed  algorithm based on the optimal structure  achieves near-optimal performance with  significantly lower computational complexity and communication overhead than existing methods. Our proposed algorithm is scalable  to the network size, in terms of  the number of BS antennas, users, and coordinating BSs, thereby enabling broader cooperation among BSs.
\end{itemize}

%%%%%%%%%%%%%%%%%%%%%%%%%%%%%%%%%%%
\subsection{Organization and Notations} \label{sec:O&N}
 The rest of this paper is organized as follows. Section~\ref{sec:ncmodel} introduces the system model and the problem formulation. In Section~\ref{OptimalStructureQoS}, we derive the optimal coordinated multicast beamforming structure. In Section~\ref{sec:AlgForQoS}, based on the optimal structure, we present our fast  computational algorithms for determining unknown parameters  and propose a semi-distributed computing approach to implement the proposed algorithm between the BSs and the CPU. In Section~\ref{sec:others}, we extend our results to the coordination design under imperfect CSI and  other coordination scenarios. The simulation results and discussion are presented in Section~\ref{sec:sim}, followed by the conclusion in Section~\ref{sec:conclusion}. 

\emph{Notations:} Hermitian, transpose, trace, and conjugate of $\Abf$ are denoted by $\Abf^{H}$, $\Abf^{T}$, $\tr(\Abf)$ and $\Abf^*$ respectively. An identity matrix is denoted by $\Ibf$. A semi-definite matrix $\Abf$ is denoted as $\Abf \succcurlyeq \mathbf{0}$. The Euclidean norm of  vector $\abf$ is denoted by $\|\abf\|$. Notation $\xbf \sim \Cc\Nc(\abf, \Cbf)$ means random vector $\xbf$ follows a complex Gaussian distribution with mean $\abf$ and covariance matrix $\Cbf$.  The abbreviation i.i.d. stands for independent and identically distributed.

\allowdisplaybreaks

%%%%%%%%%%%%%%%%%%%%%%%%%%%%%%%%%%%%%%%%%%%%%%%%%%%%%%%%%%%%%%%%%%%
\section{System Model and Problem Formulation} \label{sec:ncmodel}
We consider a downlink multicast transmission scenario in a multi-cell massive MIMO system consisting of $J$ cells, where the base station (BS) in each cell  provides the multicast service to  a group of $K$ users in its cell.\footnote{We assume one group per cell for the ease of exposition. The results obtained can be extended to the scenario with multiple groups per cell, see Section~\ref{subsec:others}. }  Each BS is equipped with $M$ antennas, and each  user is equipped with a single antenna. We assume that all BSs  use the same spectrum bandwidth for transmission. 

We assume no data sharing among BSs. Coordination among $J$ BSs is considered for  inter-cell interference management, and we study the design of coordinated multicast beamforming  among these BSs for their  multicast services. Each BS multicasts a message to the $K$ users in its own cell, using the beamforming vector that is jointly designed among all the BSs. Define the cell index set $\Jc\triangleq\{1,\cdots,J\}$ and the user index set $\Kc\triangleq\{1,\cdots,K\}$.  The serving BS in cell $j$ is denoted by BS $j$. Let ${\hbf}_{j,ik}$ denote the $M \times 1$ channel vector from BS $j$ to user $k$ in cell $i$, for  $k\in\Kc$, $j,i\in\Jc$.  Let $\wbf_i$ denote the $M \times 1$ multicast beamforming vector at BS $i$.  The received signal at user $k$ in  cell $i$ is given by
\begin{align}\label{eqn:system function}
y_{ik}=\wbf_{i}^H\hbf_{i,ik}s_{i}+\hspace*{-.2em}\sum^{J}_{\substack{j=1\\j \neq i}}\wbf_j^H\hbf_{j,ik}s_{j}+n_{ik}, \ k\in\!\Kc, i\in\!\Jc.
\end{align}
where $s_{i}$ is the data symbol transmitted from BS $i$ with ${\rm E}[|s_{i}|^{2}]=1$, and $n_{ik}$ is the receiver additive white Gaussian noise at the user with zero mean and variance $\sigma^2$. The first term in \eqref{eqn:system function} is the desired signal,  and the second term is the interference from the other BSs of the coordinated neighboring cells.
The transmit power at BS $i$ is given by $\|\wbf_{i}\|^{2}$, for $i \in \Jc$.

From (\ref{eqn:system function}), the received SINR at user $k$ in cell $i$ is given by
\begin{align}\label{eqn:SINRexpression}
\hspace*{-.4em}\SINR_{ik}
=\frac{|{\hbf}_{i,ik}^{H}{\wbf}_{i}|^{2}}{\sum^{J}_{j=1,j\neq i}|{\hbf}_{j,ik}^{H}{\wbf}_{j}|^2+\sigma^2}\ ,\  k\in\Kc,\ i\in\Jc.
\end{align}

For the coordinated multicast beamforming design, we consider the QoS problem to minimize each BS transmit power  while meeting the minimum  SINR targets of all users. For a multi-cell system, each BS may have its individual power budget, denoted by $p_{i}$, $i\in\Jc$. Thus, one way to consider such a QoS problem is to consider the transmit power margin  (w.r.t. its power budget) $\|\wbf_i\|^2/ p_{i}$ at each BS $i$  and   formulate the problem  to minimize the maximum transmit power margin of all the BSs in the coordinated cells,\footnote{Note that power budget $p_i$ at BS $i$ is not a strict power limit, but an estimate of  the BS desired power target. Depending on the SINR target $\gamma_{ik}$ and $p_i$ settings, the actual transmit power may exceed $p_i$. The objective is to minimize each BS power usage,  such that the power consumption against  its budget is minimized.} given by
\begin{align} %\label{eqn:QoSproblem0}
&\Pc_{o}: \min_{\Wbf} \max_{i } \ \frac{1}{p_i}\|\wbf_{i}\|^{2}\nn \\
&\st\  \frac{|\hbf^{H}_{i,ik}\wbf_{i}|^{2}}{\sum^{J}_{j=1,j\neq i}| \hbf^{H}_{j,ik}\wbf_{j}|^{2}+\sigma^{2}}\geqslant\gamma_{ik}, \ k \in \Kc, i \in \Jc \nn 
\end{align}
where $\Wbf\triangleq[\wbf_{1}, \cdots ,\wbf_{J}]$ is the beamforming matrix containing the  multicast beamforming vectors of all BSs, and $\gamma_{ik}$ is the minimum SINR target at user $k$ in cell $i$.

Problem $\Pc_o$ is a non-convex and NP-hard problem due to the multicast nature. Moreover, it is a large-scale optimization problem with the number of transmit antennas $M\gg 1$ in a massive MIMO system. These characteristics impose significant challenges in designing a solution that is not only of good performance but also  scalable and computationally efficient. To address these challenges, we first derive the optimal  beamforming structure for the coordinated multi-cell multicasting. Based on this  structure, we then develop a fast algorithm to obtain the solution that can be computed semi-distributively.

\section{Optimal Structure of Coordinated Multicast Beamforming}\label{OptimalStructureQoS}
Problem $\Pc_{o}$ for multi-cell coordinated multicast beamforming is a min-max optimization problem under the  individual BS transmit power budget,  which is a more difficult problem than the total BS transmit power minimization in the single-cell case   \cite{Dong&Wang:SP20}. Despite of this, we will show that we can extend the technique in \cite{Dong&Wang:SP20}   to the multi-cell scenario and derive the structure of the optimal solution to $\Pc_o$.

Using the auxiliary variable $t$, we first convert $\Pc_o$ into the following equivalent problem for $(\Wbf,t)$:    
\begin{align}
&\Pc_1: \ \min_{\Wbf,t} \  t\nn \\
&\st\ \  \frac{|\hbf^{H}_{i,ik}\wbf_{i}|^{2}}{\sum^{J}_{j=1,j\neq i}| \hbf^{H}_{j,ik}\wbf_{j}|^{2}+\sigma^{2}}\geqslant\gamma_{ik}, \  k \in \Kc,i \in \Jc  \label{eqn:QoSequalProblem1}\\
&\qquad \frac{1}{p_i} \|\wbf_{i}\|^{2}-t\leqslant0, \  i \in \Jc\label{eqn:Pwr_constr}
\end{align}
where  the constraint \eqref{eqn:Pwr_constr} is for the per-BS transmit power.
  Consider using the SCA method to iteratively solves a sequence of convex approximations of $\Pc_1$ to obtain a stationary solution. We will analyze the solution under the SCA method to derive the structure of the optimal beamforming solution to $\Pc_1$. 

\subsection{The Optimal Solution to SCA Subproblem} \label{subsec:SCA_sol}
By the SCA method, we introduce the $M \times 1$ auxiliary vector $\zbf_{i}, i \in \Jc$, and have the following inequality  for any $\wbf_{i}$, $\zbf_i$:  
\begin{align*}%\label{ineq_SCA}
\wbf^{H}_{i} \hbf_{i,ik}\hbf^{H}_{i,ik}\wbf_{i} \geqslant2
 \mathfrak{Re}\{\wbf^{H}_{i}\hbf_{i,ik}\hbf^{H}_{i,ik}\zbf_{i}\}-
 \zbf^{H}_{i}\hbf_{i,ik}\hbf^{H}_{i,ik}\zbf_{i},
\end{align*}
where the equality holds if and only if $\wbf_{i}=\zbf_{i}$.
Applying the above inequality to  the numerator
of the SINR expression in the constraint in  \eqref{eqn:QoSequalProblem1}, we obtain a lower bound on the SINR. Replacing the SINR with this lower bound,  and letting   $\Zbf \triangleq[\zbf_{1}, \ldots, \zbf_{J}]$, we obtain the following  convex approximation  of $\Pc_1$ for given $\Zbf$: 
\begin{align} 
\Pc_{1\SCA}& (\Zbf):  \ \min_{\Wbf,t} \ t \nn\\
 \st&  \ \ \gamma_{ik}\!\!\!\!\sum^{J}_{j=1,j \neq i}| \hbf^H_{j,ik}\wbf_j|^{2}-2\mathfrak{Re} \{\wbf_i^H\hbf_{i,ik}\hbf^{H}_{i,ik}\zbf_{i}\} \nn \\
&\quad \  +|\zbf_i^H\hbf_{i,ik}|^{2}
+\gamma_{ik}\sigma^{2}\leqslant0, \ k \in \Kc, i \in \Jc\label{eqn:QoSp1SCA}\\
&\ \ \frac{1}{p_i}\|\wbf_{i}\|^{2}-t\leqslant0, \ i \in \Jc \label{eqn:QoSp1SCA_pwr}
\end{align}
where the non-convex SINR constraint in \eqref{eqn:QoSequalProblem1} is replaced by the convex constraint function in \eqref{eqn:QoSp1SCA}. Let $(\Wbf^{\star}(\Zbf),t^\star(\Zbf))$ be the 
optimal solution to $\Pc_{1\SCA}(\Zbf)$, which is also is feasible to $\Pc_1$.  Replacing $\Zbf$ with the optimal solution $\Wbf^{\star}(\Zbf)$, we iteratively solve a sequence of such SCA subproblems until convergence.
This SCA method is guaranteed to converge to a stationary point $\Wbf^{\star}$ of $\Pc_1$  \cite{Marks&Wright:OR78}.  

% Hence, the above procedure is possible to converge to the global optimal solution $\wbf^{o}$ of $\Pc_{o}$, due to the fact that the global optimal solution is a stationary point. This convergence can arise if the initial point $\zbf^{\left(0\right)}$ is chosen properly, \textit{e.g.,} $\zbf^{\left(0\right)}$ is at the vicinity of $\wbf^{o}$. In that case, we have $\zbf^{\left( l \right)}\rightarrow\mathbf{w}^{\star}=\mathbf{w}^{o}$, and $t^{\left(l\right)\star}\rightarrow t^{\star}\left( \mathbf{w}^{o}\right)=t^{o}$.

%%%%%%%%%%%%%%%%%%%%%%%%%%%%%
%\subsection{Optimal Structure of Coordinated Multicast Beamforming}\label{OptimalStructure}
Since each SCA subproblem $\Pc_{1\SCA}(\Zbf)$ is  a jointly convex problem with respect to (\wrt) $(\Wbf,t)$, and Slater's condition holds, we  can obtain its optimal solution  by solving its Lagrange dual problem \cite{book:Boyd&Vandenberghe}.  The Lagrangian for $\Pc_{1\SCA}(\Zbf)$ is given by
\begin{align} \label{eqn:QoSLagrangian}
&\Lc(\Wbf,t,\lambdabf,\mubf; \Zbf) \nn\\
 &= t+\sum^{J}_{i=1}
 \mu_i\!\bigg(\!\frac{\|\wbf_{i}\|^{2}}{p_i}-t \! \bigg)\!+\!\sum^{J}_{i=1}\sum^{K}_{k=1}\lambda_{ik}
 \bigg[\gamma_{ik}\!\!\!\sum_{j=1,j \neq i}^J\!\!\big|\wbf^{H}_{j}\hbf_{j,ik}\big|^{2} \nn\\
&\ \ -2\mathfrak{Re}\{\wbf^{H}_{i}\hbf_{i,ik}\hbf^{H}_{i,ik}\zbf_{i}\}
+|\zbf^{H}_{i}\hbf_{i,ik}|^{2}+\gamma_{ik}\sigma^{2}\bigg] 
\end{align}
where $\lambda_{ik}$ and  $\mu_{i}$ are the Lagrange multipliers associated with the  QoS constraint for user $k$ in cell $i$ in \eqref{eqn:QoSp1SCA} and   BS $i$'s transmit power constraint in \eqref{eqn:QoSp1SCA_pwr}, respectively, and we denote $\lambdabf \triangleq [\lambdabf^{T}_{1}, \dots, \lambdabf^{T}_{J}]^{T}$  with $\lambdabf_{i}\triangleq[\lambda_{i1},\ldots,\lambda_{iK}]^{T}$ and $\mubf \triangleq [\mu_{1},\ldots,\mu_{J}]^{T}$. After regrouping the terms \wrt\  $t$ and $\wbf_i$ in \eqref{eqn:QoSLagrangian}, we rewrite the Lagrangian as
\begin{align} \label{eqn:QoSLagrangianRegrouping}
\Lc&(\Wbf,t,\lambdabf,\mubf; \Zbf) \nn\\
&=(1-\mathbf{1}^T\mubf)t +\sum^{J}_{i=1}\sum^{K}_{k=1}\lambda_{ik}
\left(\sigma^{2}\gamma_{ik}+|\zbf^{H}_{i}\hbf_{i,ik}|^{2}\right) \nn\\
 &\quad +\sum^{J}_{i=1}\wbf_{i}^{H} \bigg(\frac{\mu_{i}}{p_i}\Ibf+\!\!\!\sum_{j=1,j \neq i}^J\sum^{K}_{k=1}\lambda_{jk}\gamma_{jk}\hbf_{i,jk}\hbf^{H}_{i,jk}\bigg)\wbf_{i} \nn \\
&\quad -2\sum^{J}_{i=1}\mathfrak{Re}\bigg\{\zbf^{H}_{i}\bigg(\sum^{K}_{k=1}
\lambda_{ik}\hbf_{i,ik}\hbf^{H}_{i,ik}\bigg)\wbf_{i}\bigg\} \nn\\
&= (1-\mathbf{1}^T\mubf)t +\sum^{J}_{i=1}\sum^{K}_{k=1}\lambda_{ik}
\left(\sigma^{2}\gamma_{ik}+|\zbf^{H}_{i}\hbf_{i,ik}|^{2}\right) \nn\\
 &\quad +\sum^{J}_{i=1}\wbf_{i}^{H} \Rbf_{i,i^{-}}(\lambdabf,\mubf)\wbf_{i} -2\sum^{J}_{i=1}\mathfrak{Re}\big\{\nubf_i^H\wbf_i\big\}
\end{align}
where
\begin{align}
\label{eqn:QoSOptStrR-}
\Rbf_{i,i^{-}}(\lambdabf,\mu_{i}) &\triangleq\frac{\mu_{i}}{p_i}\Ibf
+\!\!\sum^{J}_{j=1,j \neq i}\sum^{K}_{k=1}\lambda_{jk}
\gamma_{jk}\hbf_{i,jk}\hbf^{H}_{i,jk}, \\
\label{eqn:QoSOptStrNu}
\nubf_i &\triangleq\bigg(\sum^{K}_{k=1}
\lambda_{ik}\hbf_{i,ik}\hbf^{H}_{i,ik}\bigg)\zbf_{i}.
\end{align}
We note that $\Rbf_{i,i^{-}}(\lambdabf,\mu_{i})$ for BS $i$ contains the sample covariance matrix of channels from BS $i$ to all users in other cells and is parameterized by both $\lambdabf$ and $\mubf$. 

The  Lagrange dual function for  $\Pc_{1\SCA}(\Zbf)$ is given by  
\begin{align}\label{eqn:QoSproblemgFunc}
g(\lambdabf,\mubf;\Zbf)\triangleq \min_{\Wbf,t }\Lc(\Wbf,t,\lambdabf,\mubf; \Zbf),
\end{align}
and the dual problem is 
\begin{align} %\label{eqn:QoSproblemD1}
\Dc_{1\SCA}(\Zbf): &\max_{\lambdabf\succcurlyeq \mathbf{0} , \mubf \succcurlyeq \mathbf{0,}}g(\lambdabf,\mubf;\Zbf). \nn 
%&\st\  \lambdabf\succcurlyeq \mathbf{0} , \mubf \succcurlyeq \mathbf{0,}
\end{align}

Solving the minimization problem in \eqref{eqn:QoSproblemgFunc} under the optimal Lagrange multipliers, we obtain the  solution
$\wbf^{\star}_{i}(\zbf)$, $i\in \Jc$, to $\Pc_{1\SCA}(\Zbf)$.  Let $\Hbf_{i}\triangleq[\hbf_{i,i1},\ldots,\hbf_{i,iK}]$ denote the channel matrix between BS $i$ and its own $K$ users in  cell $i$. The solution to $\Pc_{1\SCA}(\Zbf)$ is given in closed-form as follows.
\begin{lemma}\label{prop1}
The optimal solution $\wbf^{\star}_{i}(\Zbf)$ to $\Pc_{1\SCA}(\Zbf)$ is
\begin{align}
\wbf^{\star}_{i}(\Zbf)=\Rbf^{-1}_{i,i^{-}}(\lambdabf^{\star},\mu_i^{\star})\Hbf_i\alphabf_i^\star, \ i \in \Jc \label{eqn:OptStruc1QoS}
\end{align}
where $\lambdabf^{\star}$ and $\mubf^{\star}$ are the optimal Lagrange multipliers to the dual problem  $\Dc_{1\SCA}(\Zbf)$ satisfying  $\mathbf{1}^T\mubf^\star=1$, and $ \alphabf^{\star}_{i}\triangleq[\alpha^{\star}_{i1},\ldots,\alpha^{\star}_{iK}]^{T}$ with $\alpha_{ik}^\star \triangleq \lambda_{ik}^\star\hbf^{H}_{i,ik}\zbf_i$, $k\in\Kc, i\in \Jc$. 
\end{lemma}
\IEEEproof See Appendix~\ref{app_prop1}.

%%%%%%%%%%%%%%%%%%%%%%%%%%%%%%
\subsection{The Structure of the Optimal  Solution to $\Pc_o$}
The SCA method iteratively solve a sequence of SCA subproblems  $\Pc_{1\SCA}(\Zbf)$ by replacing $\Zbf$ with $\Wbf^{\star}(\Zbf)$ obtained from the  same subproblem in the previous iteration, until  $\Zbf$ converges to a stationary point $\wbf^{\star}$ of $\Pc_1$. If this stationary point is the global optimal solution, \ie  $\Wbf^{\star}=\Wbf^{o}$, then  $\Zbf \rightarrow \Wbf^{o}$. At the same time, the structure of $\wbf_i^\star(\zbf)$ remains as in \eqref{eqn:OptStruc1QoS}, while $\wbf_i^\star(\Zbf)$ depends on $\Zbf$ only through the optimal $(\lambdabf^{\star},\mubf^\star)$ to $\Dc_{1\SCA}(\Zbf)$ and $\alphabf_i^\star$. Following this,  the structure of the solution is stated in the following theorem.

\begin{theorem}\label{theorem-1QoS} 
The optimal  solution to the  QoS problem $\Pc_{o}$ for multi-cell coordinated multicast beamforming is given by
\begin{align}\label{eqn:thm1_w_opt}
\wbf^{o}_{i} %&= \Rbf^{-1}_{i}\left( \lambdabf^{o}, \mubf^{o}\right)\sum^{K}_{k=1}a^{o}_{ik}\hbf_{i,ik}\nn \\
&=\Rbf^{-1}_{i}( \lambdabf^{o}, \mu_i^{o})\Hbf_{i}\abf^{o}_{i},\ i \in \Jc
\end{align}
where 
\begin{align}
\label{eqn:covarianceMatrixRwithMuQoS}
\Rbf_i\left( \lambdabf, \mu_i\right)\triangleq \frac{\mu_{i}}{p_i}\Ibf+\sum^{J}_{j= 1}\sum^{K}_{k=1}\lambda_{jk}\gamma_{jk}\hbf_{i,jk}\hbf^{H}_{i,jk},
\end{align}
and $\lambdabf^{o}$ and $\mubf^{o}$ are the optimal dual solutions to $\Dc_{1\SCA}(\Wbf^o)$ with $\mathbf{1}^T\mubf^o=1$;   $\abf^{o}_{i} \triangleq [a^{o}_{i1}, \ldots, a^{o}_{iK}]^T$ contains the optimal weights of the serving users of BS $i$, with  the weight of user $k$ being  $a^{o}_{ik} = \lambda^{o}_{ik}(1+\gamma_{ik})(\hbf^H_{i,ik}\wbf^{o}_{i}),$  $k \in \Kc$, $i \in \Jc$.

The optimal objective value of $\Pc_o$ is given by 
\begin{align}\label{QoS:opt_value}
\max_{i} \ \frac{1}{p_i}\|\wbf_{i}^o\|^{2}=\sigma^{2}\lambdabf^{oT}\gammabf
\end{align}
where $\gammabf$ is the vector containing the SINR targets of all users of the $J$ coordinated cells:   $\gammabf\triangleq[\gammabf^{T}_{1},\ldots, \gammabf^{T}_{J}]^{T}$  with $\gammabf_{i}\triangleq[\gamma_{i1},\ldots\gamma_{iK}]^{T}$, $i\in\Jc$.
\end{theorem}
\IEEEproof See Appendix~\ref{appA}. 
\endIEEEproof

\begin{remark}\label{remark1}
The  optimal coordinated multicast beamformer $\wbf^{o}_{i}$ for  BS $i$  in \eqref{eqn:thm1_w_opt} is essentially a weighted MMSE beamformer. The matrix $\Rbf_i\left( \lambdabf, \mu_i\right)$ in \eqref{eqn:covarianceMatrixRwithMuQoS} is a noise-plus-weighted-channel-covariance matrix for BS $i$. Its first term is   the normalized receiver noise power scaled by $\mu_i/p_i$. Since $\mu_i$ is the Lagrange multiplier associated with BS $i$'s transmit power constraint  in \eqref{eqn:Pwr_constr},  it  can be viewed as a weight  to BS $i$'s  power budget  $p_i$. The second term contains the channels from BS $i$ to all users in $J$ cells $\{\hbf_{i,jk}, k\in\Kc,j\in\Jc\}$. We notice that the relative weight of each user channel is determined by $\lambda_{jk}\gamma_{jk}$, for user $k$ in cell $j$, which  is user specific and is the same in all  $\Rbf_i\left( \lambdabf, \mu_i\right)$'s.
The term $\Hbf_{i}\abf^{o}_{i}$ is the weighted sum of the serving user channels in cell $i$. In particular, $\hat{\hbf}_i \triangleq \Hbf_{i}\abf^{o}_{i}$ acts as the {group-channel direction} of the user group, where the optimal weight vector $\abf_i^o$ indicates the relative significance of each user  channel in this group-channel direction. It determines the beamformer $\wbf_i^o$. Thus, the optimal structure shows that even though the dimension of $\wbf_i$ may be high for large $M$,  the unknown variables are only in  $\abf_i$, which is a $K\times 1$ vector  in the user dimension. This inherent low-dimensional
structure is the key for devising a highly efficient computational method to determine $\wbf_i$.
\end{remark}

\begin{remark}\label{remark2}
We note that for the multi-cell scenario,  the optimal  $\wbf^{o}_i$ for BS $i$ in \eqref{eqn:thm1_w_opt} is only a function of the  channels from BS $i$ to all users in $J$ cells $\{\hbf_{i,jk}, , \forall k,j\}$, \ie the local CSI. Therefore, structure-wise, the optimal coordinated multicast beamformers $\{\wbf^{o}_1,\ldots,\wbf^{o}_J\}$ are naturally \emph{distributed} beamformers: each beamformer $\wbf^{o}_i$ can be computed locally at BS $i$ using local CSI without requiring the knowledge of global CSI from other cells. This inherent property is highly desirable for multi-cell coordination, as it reduces the required fronthaul communication  among the coordinating BSs. At the same time, determining the parameters in  $\wbf_i^o$ requires information exchange among BSs. In particular, we note that the optimal solution $\wbf^{o}_{i}$ in \eqref{eqn:thm1_w_opt}  is shown in a semi-closed-form, where $\lambdabf^{o}$, $\mu_i^{o}$, and $\abf^{o}_{i}$ need to be computed, and determining their optimal values requires considering $J$ cells jointly.   
\end{remark}

\begin{remark}\label{remark3}
We point out the differences of the optimal structure in \eqref{eqn:thm1_w_opt} for the coordinated BSs in the multi-cell case from that of the multi-group multicast beamforming in the single-cell case in \cite{Dong&Wang:SP20}: First, the covariance matrix  $\Rbf_i\left( \lambdabf, \mu_i\right)$ in \eqref{eqn:thm1_w_opt} contains additional parameter $\mu_i$ as the result of individual  BS transmit powers, and it  depends on power budget $p_i$. Second,  $\Rbf_i\left( \lambdabf, \mu_i\right)$ is specific to each BS $i$, which contains the channels from BS $i$ to all users in $J$ cells $\{\hbf_{i,jk}, \forall k,j\}$. This is different from the single-cell case, where a common covariance matrix is shared among all multicast beamformers. However, we note that although $\Rbf_i\left( \lambdabf, \mu_i\right)$ is different  for each BS $i$,  $\lambdabf$ is common for all  $\Rbf_i\left( \lambdabf, \mu_i\right)$'s. 
\end{remark}

%%%%%%%%%%%%%%%%%%%%%%%%%%%%%%%%%%%%%%%%%%%%%%%%%%
\section{ Fast Algorithms with Semi-Distributed Computing} \label{sec:AlgForQoS}
As discussed in Remark~\ref{remark2},   fully determining the optimal  $\wbf^{o}_{i}$  in  \eqref{eqn:thm1_w_opt} requires obtaining the parameters  $\{\lambdabf^{o}, \mu_i^{o}\}$ and weight vector $\abf^{o}_{i}$. However, finding the optimal  $\lambdabf^{o}, \mubf^{o}$ and $\{\abf_i^{o}\}$ is difficult, since $\Pc_1$ is an NP-hard problem. Thus, we need to devise effective algorithms to compute them suboptimally. Furthermore, although optimizing   $\lambdabf$, $\mubf$, and $\{\abf_i\}$ requires considering $J$ coordinating cells jointly, it is still desirable to develop a method to compute them in a distributed manner, which is also computationally efficient.     Aiming at this goal, below, we develop  semi-distributed  fast  algorithms to  compute their values. 

%%%%%%%%%%%%%%%%%%%%%%%%%%%%%%%%%%%%%%%%%%%%%%%%%%
\subsection{Computing $\Rbf_i\left( \lambdabf, \mu_i\right)$}\label{sec:Lambda&Mu}

We need to determine $\{\lambdabf, \mu_i\}$ to compute $\Rbf_i\left( \lambdabf, \mu_i\right)$ at each BS $i$. We first examine  the optimal $\lambdabf^o$ and $\mubf^o$ in the optimal solution $\wbf_i^o$ in \eqref{eqn:thm1_w_opt}.  From Theorem~\ref{theorem-1QoS}, we have $a^{o}_{ik} \triangleq \lambda^{o}_{ik}(1+\gamma_{ik})(\hbf^{H}_{i,ik}\wbf^{o}_{i})$, $\forall k,i$. Let $\delta_{ik} \triangleq \hbf^{H}_{i,ik}\wbf_i^{o}$, $k\in \Kc$, and $\deltabf_i = [\delta_{i1},\ldots,\delta_{iK}]^T=\Hbf_i^H\wbf_i^o$, $i\in\Jc$. Also, let $\Dbf_{ \lambdabf_{i}} \triangleq \text{diag}(\lambdabf_i)$. Then, we can express $a^{o}_{ik}$ into the vector form as
$\abf^{o}_{i} = \Dbf_{ \lambdabf_{i}}\left( \Ibf + \Dbf_{\gammabf_{i}} \right)\deltabf_{i}$, $i\in \Jc$.
Based on the optimal solution in \eqref{eqn:thm1_w_opt},  we have  
\begin{align}
\label{eqn:deltaMatDefQoS}%\label{eqn:deltaDefQoS}
\deltabf_{i}  &= \Hbf^{H}_{i}   \Rbf^{-1}_{i}(  \lambdabf^{o},  \mu_i^{o}) \Hbf_{i}  \abf_{i}^o \nn \\
& = \Hbf^{H}_{i} \Rbf^{-1}_{i}(  \lambdabf^{o},  \mu_i^{o}) \Hbf_{i} \Dbf_{ \lambdabf_{i}}( \Ibf + \Dbf_{\gammabf_{i}} )\deltabf_{i},
\end{align}
which leads to
\begin{align} \label{eqn:lambdaSatisfyQoS}
\left( \Hbf^{H}_{i} \Rbf^{-1}_{i}( \lambdabf^{o},  \mu_i^{o}) 
\Hbf_{i} \Dbf_{\lambdabf_{i}}
\left( \Ibf + \Dbf_{\gammabf_{i}} \right) - \Ibf \right) 
\deltabf_{i} = \mathbf{0}.
\end{align}
Thus, at the optimality, the optimal $(\lambdabf^{o}, \mu_i^{o})$ should satisfy \eqref{eqn:lambdaSatisfyQoS}, for any $i\in \Jc$. However, with unknown $\deltabf_i$, it is difficult to find $(\lambdabf^{o},\mu_i^o)$ based on  \eqref{eqn:lambdaSatisfyQoS}.  
One way is to consider a sufficient condition for \eqref{eqn:lambdaSatisfyQoS},  given by
 \begin{align}
\Hbf^{H}_{i} \Rbf^{-1}_{i}( \lambdabf^{o}, \mu_i^o) 
\Hbf_{i} \Dbf_{ \lambdabf_{i}}
\left( \Ibf + \Dbf_{\gammabf_{i}} \right) = \Ibf, \ \ i \in \Jc,
\label{eqn:SuffCondRi}
\end{align}
which can be described  element-wise as follows for $i\in \Jc$:
\begin{align}\label{eqn:CondiForLambdaQoS}
\hspace*{-.55em}\begin{cases}
\lambda_{ik}\left( 1+\gamma_{ik}\right)
\hbf^{H}_{i,ik} \Rbf^{-1}_{i}( \lambdabf, \mu_i) 
\hbf_{i,ik}=1, & \!\! k \in \Kc \\
\lambda_{ik}\left( 1+\gamma_{ik}\right)
\hbf^{H}_{i,ik} \Rbf^{-1}_{i}(  \lambdabf, \mu_i) 
\hbf_{i,il}=0, & \!\! l \neq k, l \in \Kc. \\
\end{cases}
\end{align}
Note that although equations in \eqref{eqn:CondiForLambdaQoS} are functions of $\mu_i$, $\lambdabf$ is common for  all $i\in \Jc$.         
Assuming  $\mubf$ is given,  we  note that  \eqref{eqn:CondiForLambdaQoS} as a sufficient condition, typically contains more equations than variables, and thus,   $\lambda_{ik}$ may not satisfy all the equations. We propose to compute $\lambdabf$ using a method similar to the one proposed in \cite{Dong&Wang:SP20}. That is, we
consider the first equation  in \eqref{eqn:CondiForLambdaQoS} only  (\ie the diagonal elements of the matrix equation in \eqref{eqn:SuffCondRi}) and solve  $\lambdabf$  using the fixed-point iterative method:
\begin{align}\label{lambda:fixed-point}
\lambda^{(m+1)}_{ik}=
\frac{1}{\left( 1+\gamma_{ik}\right)
\hbf^{H}_{i,ik}\Rbf^{-1}_{i}
\big( \lambdabf^{(m)},\mu_i\big) \hbf_{i,ik}}, \ \forall k,i.
\end{align}
where $m$ is the iteration index. The detail of the algorithm will be described at the end of this subsection when we discuss the semi-distributed implementation. 
 
\begin{remark}
Although we only used the first equation in  \eqref{eqn:CondiForLambdaQoS} to compute $\lambdabf$, we expect that for massive MIMO with  $M$ being large, the second equation can be approximately satisfied. To see this, we can interpret the expression at the left-hand-side as the  channel correlation of two users  $k$ and $l$ in serving cell $i$ defined by  $\Rbf^{-1}_{i}(  \lambdabf, \mu_i)$. Since the two user channels are typically independent to each other and with zero-mean elements, we expect the channel correlation w.r.t.   $\Rbf^{-1}_{i}(  \lambdabf, \mu_i)$ goes to $0$ as $M \to \infty$, and  $\lambdabf$  computed by  \eqref{lambda:fixed-point} asymptotically satisfies \eqref{eqn:SuffCondRi}.
\end{remark}
   
 For determining $\mubf$, Theorem~\ref{theorem-1QoS} shows that $\mathbf{1}^H\mubf^o=1$.  However,  it is difficult to find  the  values of $\mu_i$'s. Note from Remark~\ref{remark1} that,   $\mu_i$ acts as a weight  in  $\Rbf_i\left( \lambdabf, \mu_i\right)$ in \eqref{eqn:covarianceMatrixRwithMuQoS} for  the power budget $p_i$ at BS $i$. To avoid over-complicated  computation, we propose to  uniformly set $\mu_i = 1/J$, $\forall i \in \Jc$.
  In the  case when all BSs have the same power budget, $p_i = p$, $\forall i$, we expect all BSs are weighted equally, and  each BS on average has the  similar  transmit power margin over its power budget. 
Thus, we set $\mu_i = 1/J$ in $\Rbf_i\left( \lambdabf, \mu_i\right)$'s for the rest of the computation. We will see in the simulation results that in the case of $p_i=p$, $\forall i$,  our proposed approach is effective and provides a near-optimal performance.

\subsubsection{Semi-Distributed Implementation} \label{subsubsec:R}
 The above proposed method for computing $\lambdabf$ and thus $\Rbf_i\left( \lambdabf, \mu_i\right)$ can be implemented in a semi-distributed manner at each BS. To see this, note from \eqref{lambda:fixed-point} that computing each element $\lambda_{ik}^{(m+1)}$ in $\lambdabf_i^{(m+1)}$ only requires channels $\{\hbf_{i,jk},\forall k,j\}$ available at BS $i$ and $\lambdabf^{(m)}$ from previous iteration. Thus,  BSs only need to exchange   $\lambdabf_i^{(m)}$'s from the previous iteration to update $\Rbf_i( \lambdabf^{(m)}, \mu_i)$, and   $\lambdabf_i^{(m+1)}$ can be computed distributively at each BS $i$. The required information exchange per iteration is $\lambdabf^{(m)}$ with  $JK$  real-valued elements, which is independent of $M$.
This semi-distributed  method is shown in 
 Algorithm \ref{alg:LambdaObtainQoS}. Since the method only uses a closed-form update, it is computationally efficient.

%%%%%%%%%%%%%%%%%%%%%%%%%%%%%%%%%%%%%%%%%%%%%%%%%%
\begin{algorithm}[t]
\begin{algorithmic}[1]
\caption {Semi-Distributed Method to Compute $\Rbf_i\left( \lambdabf, \mu_i\right)$}
\label{alg:LambdaObtainQoS}
\State \textbf{Initialization}: Set $\lambdabf^{(0)} \succcurlyeq \bf{0}$ for all BSs;
Set  $m=0$.  
\Repeat
\State
{\bf At each BS $i \in \Jc$:}
\State \quad Compute $\Rbf_{i}(\lambdabf^{(m)}, \mu_i)$ using \eqref{eqn:covarianceMatrixRwithMuQoS}.
\State \quad For all  $k \in \Kc$, compute
\begin{align}
\lambda^{(m+1)}_{ik}=
\frac{1}{\left( 1+\gamma_{ik}\right)
\hbf^{H}_{i,ik}\Rbf^{-1}_{i}
\big( \lambdabf^{(m)},\mu_i\big) \hbf_{i,ik}}.\nn
\end{align}
\State \quad $m\leftarrow m+1$.
\State {\bf BSs exchange $\lambdabf_i^{(m)}$'s.}
\Until convergence %{$\max_{i,k}\left|\lambda^{(m+1)}_{ik}-
%\lambda^{(m)}_{ik} \right|\leqslant\epsilon$}.
\end{algorithmic}
\end{algorithm}

\subsection{Fast Algorithm for  Weight ${\abf_i}$}
\label{sec:ADMM-SCA-QoS}
% Given $\lambdabf^{o}$ and $\mubf$ and applying the optimal solution structure in \eqref{eqn:covarianceMatrixRwithMuQoS}, the original problem $\Pc_{o}$ is transformed into $\Pc_{2\SCA}(\ubf)$. Note that the high computational cost caused by massive MIMO system is reduced by replacing the beamformer $\wbf$ to the optimal structure with low dimension weight $\abf$,  solved iteratively by updating $\ubf$ with the solution $\abf$.
%\subsubsection{The SCA Method}\label{sec:OptSCAQoS}

Once $\Rbf_i\left( \lambdabf, \mu_i\right)$ is obtained,  only weight vector $\abf_i$ needs to be computed to determine $\wbf_i$  in  \eqref{eqn:thm1_w_opt}. Let $\abf \triangleq [\abf_1^H,\ldots,\abf_J^H]^H$ be the concatenated weight vector. Based on the optimal beamforming structure in   \eqref{eqn:thm1_w_opt}, we can convert the original problem  $\Pc_{1}$ \wrt $(\Wbf,t)$ into a joint optimization of $(\abf,t)$, given by
%\vspace*{-1.5em}
\begin{align}
&\hspace*{-.5em}\Pc_{2}: \min_{\abf,t} \quad t \nn \\
&\hspace*{-.5em}\st\  \frac{|\abf^{H}_{i}\Gbf_i^H\hbf_{i,ik}|^{2}}
{\sum^{J}_{j=1,j\neq i}|\abf^{H}_{j}\Gbf_j^H\hbf_{j,ik}|^{2}+\sigma^{2}}\geqslant\gamma_{ik}, k \in \Kc, i \in \Jc \label{eqn:P2expressByaQoS}\\
&\quad \quad\frac{1}{p_i}\|\Gbf_i\abf_i\|^{2}-t\leqslant0, \ i \in \Jc \nn
\end{align}
where $\Gbf_i \triangleq \Rbf^{-1}_{i}(\lambdabf, \mu_i)\Hbf_{i}$. 
Note that  the dimension of $\abf$ is $JK$, which is independent of $M$. Thus, by the above conversion, $\Pc_2$ has a much smaller size than $\Pc_1$ of size  $JM$ for $\Wbf$, for $K \ll M$, which is particularly beneficial for massive MIMO systems.

We  consider the SCA method to solve $\Pc_{2}$ for $\abf$  iteratively, similar to $\Pc_{1\SCA}$.
Specifically, denote $\ubf\triangleq [\ubf^{H}_{1}, \ldots, \ubf^{H}_{J}]^{H}$, where $\ubf_{i}$ is $K \times 1$ auxiliary vector for each $\abf_i$. Given $\ubf$, we apply the convex approximation to the SINR constraint in \eqref{eqn:P2expressByaQoS} and have the following joint optimization subproblem w.r.t. $(\abf, t)$ at each SCA iteration:
\begin{align}
%\label{eqn:P2SCAQoS}
\Pc_{2\SCA}(\ubf):& \min_{\abf, t}\ t \nn \\
\ \st\ & \sum^{J}_{j=1,j\neq i}|\abf^{H}_{j}\fbf_{j,ik} |^{2}-2\mathfrak{Re}\left\{\abf^{H}_{i} \fbf_{i,ik}\fbf^{H}_{i,ik} \ubf_{i} \right\} 
\nn \\
& \quad + | \ubf^{H}_{i}\fbf_{i,ik} |^{2}+\sigma^{2}  \leqslant 0,\ k \in \Kc, i\in \Jc \nn \\
& \frac{1}{p_i} \|\Gbf_{i}\abf_{i}\|^{2}-t\leqslant0, \ i \in \Jc. \nn
\end{align}
where    $\fbf_{j,ik}\triangleq \Gbf_j^H\hbf_{j,ik}$, for $k \in \Kc$, $j,i \in \Jc$. 

The iterative procedure  is the same as that described in Section~\ref{subsec:SCA_sol}: After obtaining the solution $\abf^\star(\ubf)$ to $\Pc_{2\SCA}(\ubf)$, we  update $\ubf$ as $\ubf \leftarrow \abf^\star(\ubf)$, and solve $\Pc_{2\SCA}(\ubf)$ iteratively until convergence.

Each SCA iteration needs to solve the convex  subproblem
 $\Pc_{2\SCA}(\ubf)$, which  can be computed using the interior-point algorithm \cite{book:Boyd&Vandenberghe} by the standard convex solvers. However, it requires to compute $\abf_i$'s jointly and is a centralized method for beamforming among coordinating BSs. Furthermore, it is a second-order algorithm with a relatively high computational complexity, especially when the problem size grows and the subproblem needs to be solved repeatedly in each SCA iteration, which is undesirable.   To address these issues,  we propose a    fast algorithm to compute $\abf_i$  in a semi-distributive manner at each BS $i$ efficiently.

\subsubsection{ADMM Construction for $\Pc_{2\SCA}(\ubf)$}\label{subsec:ADMM}

 We  explore ADMM technique \cite{Zhangetal&Dong:SP23} to solve $\Pc_{2\SCA}(\ubf)$ at each SCA iteration. ADMM is a robust numerical method that can provide fast computation to solve 
large-scale problems.  It can be used to break down a large problem into small subproblems to be solved individually with lower computational complexity. However, whether ADMM can be an efficient algorithm depends on the specific problem structure and the   ADMM construction for that problem. In particular, since ADMM construction is not unique, it is essential that the construction design can lead to subproblems that yield computationally efficient solutions or even closed-form solutions, and at the same time, they can be distributively computed.    

For our ADMM construction,  we  introduce the auxiliary variables   $v \in \mathbb{R}$ and  $d_{j,ik} \in \mathbb{C}$, $k \in \Kc$, $i,j \in \Jc$, and  transform $\Pc_{2\SCA}(\ubf)$ into the following equivalent problem:
\begin{align}
 \Pc_{\ADMM}&(\ubf):\ \min_{\abf, \dbf,t, v} t \nn \\
\st& \ \ d_{j,ik} = \abf^{H}_{j} \fbf_{j,ik}, \  k \in \Kc,\ i,j\in \Jc, \label{ADMM:equal_1} \\
&\ \ v=t, \label{ADMM:equal_2} \\
&\ \ \gamma_{ik}\sum^{J}_{j=1,j \neq i}|d_{j,ik}|^{2} +
|\ubf^{H}_{i} \fbf_{i,ik}|^{2}+\gamma_{ik}\sigma^{2}  \nn \\
&\quad \quad-2\mathfrak{Re}\{d_{i,ik} \fbf^{H}_{i,ik} \ubf_{i}\} 
\leqslant 0, \ k \in \Kc, i \in \Jc, \label{eqn:ADMM-inEqualConsQoS}\\
&\ \ \frac{1}{p_i}\|\Gbf_{i} \abf_{i} \|^{2} -v \leqslant 0,\ i \in \Jc \label{eqn:ADMM-indiPowerQoS}
\end{align} 
where $\dbf \triangleq [\dbf_{11}^{H}, \ldots, \dbf_{JK}^{H}]^{H} \in \mathbb{C}^{J^2K}$ with $\dbf_{ik}\triangleq [d_{1,ik},\ldots, d_{J,ik}]^{T}$.

Denote the feasible set for $\dbf$ satisfying the  constraint \eqref{eqn:ADMM-inEqualConsQoS} as $\Fc$, and that for $(\abf,v)$ satisfying the constraint  \eqref{eqn:ADMM-indiPowerQoS} as $\Cc$. Define  the indicator functions for  $\Fc$ and $\Cc$ respectively as 
\begin{align}
\label{eqn:ADMMindicatorFQoS}
I_{\Fc}(\dbf)\!\triangleq\!\begin{cases}0 & \dbf \in \Fc \\
\infty & \textrm{o.w.} \\
\end{cases}, \ \
%\end{align}
%\begin{align}
%\label{eqn:ADMMindicatorCQoS}
I_{\Cc}(\abf, v)\!\triangleq\!\begin{cases}0 & (\abf, v) \in \Cc \\
\infty & \textrm{o.w.} \\
\end{cases}.
\end{align}
Then, we can  transform $\Pc_{\ADMM}(\ubf)$ into the following equality-constrained problem: 
\begin{align}
%\label{eqn:ADMM'QoS}
\Pc_{\ADMM}^{'}(\ubf): &\min_{\abf, \dbf,t, v} \ t +I_{\Fc}(\dbf) +I_{\Cc}(\abf, v) \nn \\
\st &\ \ d_{j,ik}=\abf^{H}_{j} \fbf_{j,ik}, \ k \in \Kc,\ i,j\in \Jc \nn \\
& \ \ v=t. \nn 
\end{align}
Based on the ADMM technique, the augmented Lagrangian of $\Pc_{\ADMM}^{'}(\ubf)$ is given by 
\begin{align}
&\Lc_{\rho}(\abf, \dbf,t, v, \qbf, z)=
t+I_{\Fc}(\dbf)+I_{\Cc}(\abf, v) \label{eqn:ADMMLagrangianQoS}
 \\
& +\frac{\rho}{2}\sum_{j=1}^{J}\sum_{i=1}^{J}\sum_{k=1}^{K}
|d_{j,ik} - \abf^{H}_{j} \fbf_{j,ik} + q_{j,ik}|^{2}+\frac{\rho}{2}(v-t+z)^{2} \nn
\end{align}
where $\rho > 0$ is the penalty parameter, and  $\{q_{j,ik} \in \mathbb{C}, k \in \Kc,\ i,j\in \Jc\}$  and $z \in \mathbb{R}$ are the dual variables associated with the respective equality constraints in  $\Pc_{\ADMM}^{'}(\ubf)$. Also, we denote $\qbf \triangleq [\qbf_{11}^{H}, \ldots, \qbf_{NK}^{H}]^{H}$ with $\qbf_{ik} \triangleq [q_{1,ik}, \ldots, q_{J,ik}]^{T}$.

Note that our particular design of ADMM\ construction lies in the auxiliary variables $(\dbf, v)$ and their respective equivalency constraints in \eqref{ADMM:equal_1} and \eqref{ADMM:equal_2}. They enable us to break the minimize of $\Lc_{\rho}(\abf, \dbf,t, v, \qbf, z)$ into smaller subproblems. Specifically, we note that the terms in \eqref{eqn:ADMMLagrangianQoS} for  $(\dbf, v)$ and $(\abf, t)$ are separate.\ Thus, the optimization of  $\Lc_{\rho}(\abf, \dbf,t, v, \qbf, z)$ can be decomposed into two subproblems  for $(\dbf, v)$ and  $(\abf, t)$ separately, which can be solved alternatingly. 

The proposed ADMM-based algorithm for $\Pc_{2\SCA}(\ubf)$  is summarized below:
\begin{center} \fbox{
\begin{minipage}{.94\columnwidth}
Initialize $\qbf^{(0)}, z^{(0)},t^{(0)}$; Set $\abf^{(0)}=\ubf$. 

At iteration $l$:
\begin{enumerate}[leftmargin=*]
\item Update the auxiliary variables $\dbf^{(l+1)}$ and  $v^{(l+1)}$
    \begin{align}\label{eqn:ADMM-update-1}
    \hspace*{-2.3em}\{\dbf^{(l+1)}\!,\! v^{(l+1)}\} \!=\! \mathop{\arg\min}_{\dbf, v} \Lc_{\rho}(\abf^{(l)}\!, t^{(l)}\!, v^{(l)}\!, \dbf, \qbf^{(l)}\!, z^{(l)})
 \end{align}
\item Update weight vector  $\abf^{(l+1)}$  and objective value $t^{(l+1)}$
 \begin{align} \label{eqn:ADMM-update-2}
    \hspace*{-2.7em} \{\abf^{(l+1\!)}\!, t^{(l+1\!)}\!\} \!=\! \mathop{\arg\min}_{\abf, t}  \Lc_{\rho}(\abf, t, v^{(l+1)}\!, \dbf^{(l+1)}\!, \qbf^{(l)}\!, z^{(l)})
 \end{align}
\item Update dual variables $\qbf^{(l+1)}$ and $z^{(l+1)}$
 \begin{align}
  \hspace*{-2.7em}   q_{i,jk}^{(l+1)}&=q_{i,jk}^{(l)} + \left(d_{i,jk}^{(l+1)}-
   \abf_{i}^{(l+1)H} \fbf_{i,jk} \right), \forall i, j, k  \label{eqn:ADMM-update-3}\\
  \hspace*{-2.7em}    z^{(l+1)} &= z^{(l)}+(v^{(l+1)}-t^{(l+1)}). \label{eqn:ADMM-update-4}
\end{align}
\end{enumerate}
\end{minipage}
}
\end{center}

The above ADMM  procedure contains three updating blocks in each iteration. The first two ADMM blocks involve solving two optimization subproblems \wrt   $(\dbf, v)$ and $(\abf, t)$ in \eqref{eqn:ADMM-update-1} and \eqref{eqn:ADMM-update-2}, respectively. We will show that these subproblems yield closed-form solutions, and they can be computed semi-distributively. As a result, our specific ADMM construction  leads to a semi-distributed fast  algorithm to compute the solution for $\Pc_{2\SCA}(\ubf)$ at each SCA iteration.
Finally,
since $\Pc_{2\SCA}(\ubf)$ is convex, the above ADMM procedure is guaranteed to converge to the optimal solution of  $\Pc_{2\SCA}(\ubf)$  \cite{Boyd&etal:2011ADMM}. Thus, our proposed semi-distributive algorithm obtains the  optimal solution to  $\Pc_{2\SCA}(\ubf)$. Below, we first describe the solution to each subproblem, and in Section~\ref{subsec:semi-distr}, we present the semi-distributive implementation of the algorithm.

%%%%%%%%%%%%%%%
\subsubsection{Closed-Form $(\dbf, v)$-Update}
From the expression of $\Lc_{\rho}(\abf, \dbf,t, v, \qbf, z)$ in  \eqref{eqn:ADMMLagrangianQoS},
only the second, fourth, and fifth terms are functions of  $\dbf$ and $v$. Thus, the optimization  of $\dbf$ and $v$ in \eqref{eqn:ADMM-update-1} can be separated into the following two subproblems
\begin{align}
\label{eqn:ADMMdQoS}
 &\Pc_{\dbf}(\ubf): \min_{\dbf} \sum_{j=1}^{J}
\sum_{i=1}^{J}\sum_{k=1}^{K}\big|d_{j,ik} - \abf_{j}^{(l)H} \fbf_{j,ik} + q_{j,ik}^{(l)}\big|^{2} \nn \\
&\st\ \gamma_{ik} \sum^{J}_{j=1,j\neq i}|d_{j,ik}|^{2} + |\ubf^{H}_{i} \fbf_{i,ik} |^{2}+\gamma_{ik}\sigma^{2}  \nn \\ 
&\qquad \qquad -2\mathfrak{Re}\{ d_{i,ik} \fbf_{i,ik}^H \ubf_i\}\leqslant 0, \ \  k \in \Kc, i \in \Jc.
\end{align}
and 
\begin{align}
\label{eqn:ADMMvQoS}
\Pc_{v}: \min_{v}&\ \big(v-t^{(l)}+z^{(l)}\big)^2 \nn \\ \st &  \ \frac{1}{p_i}\|\Gbf_{i}
\abf_{i}^{(l)} \|^{2} \leqslant v,  i \in \Jc.
\end{align}

Note that subproblem  $\Pc_{\dbf}(\ubf)$ can be further decomposed into $JK$ subproblems, one for each $\dbf_{ik}\triangleq [d_{1,ik},\ldots,d_{J,ik}]^T$ for each user $k$ in cell $i$, as
\begin{align}
\Pc_{\dbf_{ik}}(\ubf): \ &\min_{\dbf_{ik}}\ \sum^{J}_{j=1} \Big|d_{j,ik}-\abf^{(l)H}_{j} \fbf_{j,ik}+q^{(l)}_{j,ik}\Big|^{2} \nn \\
\st& \ \gamma_{ik} \!\!\!\sum^{J}_{j=1,j\neq i}|d_{j,ik}|^{2} + |\ubf^{H}_{i} \fbf_{i,ik} |^{2}+\gamma_{ik}\sigma^{2}  \nn \\ 
&\quad -2\mathfrak{Re}\{ d_{i,ik} \fbf_{i,ik}^H \ubf_i\}\leqslant 0.
\label{eqn:ADMMdSubQoS}
\end{align}

Note that $\Pc_{\dbf_{ik}}(\ubf)$ is a convex QCQP-1 problem, for which a closed-form solution can be obtained via the KKT conditions \cite{book:Boyd&Vandenberghe}. The closed-form
solution for such a QCQP-1 problem has been discussed in \cite{Zhangetal&Dong:SP23}, which can be used directly. For the sake of completeness, the optimal solution is provided in \eqref{eqn:ADMM-dj,ikQoS} of Appendix~\ref{appD}. 

Subproblem $\Pc_v$ can be equivalently rewritten as
\begin{align*}
%\label{eqn:ADMMvOptProbQoS}
\Pc_{v}: \min_{v} &\   (v-\!t^{(l)}\!+z^{(l)})^2 & \nn\\
\st &\  v\geqslant\max\frac{1}{p_i}\| \Gbf_{i}\abf_{i}^{(l)}\|^2
\end{align*}
which is  a quadratic program with a linear constraint. Setting the derivative of the objective function to $0$ yields $v = t^{(l)}-z^{(l)}$.  
Thus, the optimal solution $v^{o}$ is given by
\begin{align}
v^{o}= \max \bigg\{ \max_{i}\frac{1}{p_i}\| \Gbf_{i} \abf_{i}^{(l)} \|^{2}, \ t^{(l)}-z^{(l)} \bigg\}.
\label{eqn:ADMM-vExpressQoS}
\end{align}

\subsubsection{Semi-Closed-Form $(\abf, t)$-Update}\label{subsec:ADMM_end}
From   \eqref{eqn:ADMMLagrangianQoS}, the joint optimization of $\abf$ and $t$ in \eqref{eqn:ADMM-update-2} is equivalent to  the following problem:\footnote{In \eqref{eqn:ADMM-subaQoS}, we switch the indexes $i$ and $j$ in the objective function for a more consistent presentation using $\abf_i$, which does not affect the original objective function. }
\begin{align}
\label{eqn:ADMM-subaQoS}
\min_{\abf, t} & \ t+ \frac{\rho}{2}\sum_{i=1}^{J}\sum_{j=1}^{J}\sum_{k=1}^{K}| d^{(l+1)}_{i,jk} - \abf^{H}_{i} \fbf_{i,jk} + q^{(l)}_{i,jk} |^{2}\ \\ 
&+\frac{\rho}{2} \Big(v^{(l+1)} - t+z^{(l)} \Big)^{2}  \nn\\
\st &\ \frac{1}{p_i} \| \Gbf_{i} \abf_{i} \|^{2}-v^{(l+1)}\leqslant 0,\ i \in \Jc. \nn
\end{align}
Again, the above joint optimization problem can be  decomposed into two subproblems for $t$ and $\abf$ to be solved separately. The subproblem for $t$ is given by
\begin{align}
\Pc_{t}: \ \min_{t} \ t+\frac{\rho}{2}\left( v^{(l+1)} - t+z^{(l)} \right)^{2}, \label{eqn: ADMM-Pct}
\end{align}
which is an unconstrained convex quadratic  optimization problem, whose optimal solution can be easily  obtained as  
\begin{align}\label{eqn: t_opt}
t^o=v^{(l+1)}+z^{(l)}-\frac{1}{\rho}.
\end{align}

The subproblem for $\abf$ can be further decomposed into $J$ subproblems, one for each $\abf_{i}$ as
\begin{align}
\label{eqn:ADMMpSubA-QoS}
\Pc_{\abf_i}(\ubf): \ \min_{\abf_{j}}& \ \sum_{j=1}^{J}\sum_{k=1}^{K}| d^{(l+1)}_{i,jk} - \abf^{H}_{i} \fbf_{i,jk} + q^{(l)}_{i,jk} |^{2} \nn \\
\st& \  \frac{1}{p_i} \| \Gbf_{i} \abf_{i} \|^{2} \leqslant v^{(l+1)}.
\end{align}

The above problem is again a convex QCQP-1 problem, which can be solved by  the KKT conditions. The Lagrangian of $\Pc_{\abf_{i}}(\ubf)$ is given by
\begin{align}
\Lc(\abf_{i}, \tilde{\lambda}_{i})&=\sum_{j=1}^{J}\sum_{k=1}^{K}| d^{(l+1)}_{i,jk} - \abf^{H}_{i} \fbf_{i,jk} + q^{(l)}_{i,jk} |^{2}  \nn \\
&\qquad + \tilde{\lambda}_i \left( \frac{1}{p_i}\| \Gbf_i \abf_i \|^{2}-v^{(l+1)} \right)
\label{eqn:ADMM-LagraSubQoS}
\end{align}
where $\tilde{\lambda}_i$ is the Lagrangian multiplier associated with the constraint in \eqref{eqn:ADMMpSubA-QoS}. Setting  $\nabla_{\abf_i} \Lc(\abf_i, \tilde{\lambda}_i)=0$, we obtain the optimal $\abf_i$ as 
\begin{align}\label{eqn:ajFromSubAQoS}
\abf_i = \left( \frac{\tilde{\lambda}_i}{p_i}
\Gbf^{H}_i\Gbf_i+\sum_{j=1}^{J}\sum_{k=1}^{K} \fbf_{i,jk}\fbf_{i,jk}^{H} \right)^{-1} \nn \\
\cdot \sum_{j=1}^{J}\sum_{k=1}^{K}\Big( d^{(l+1)}_{i,jk} +q_{i,jk}^{(l)}\Big)^{*}\fbf_{i,jk}.
\end{align}

The optimal $\tilde{\lambda}^o_{j}$ can be determined using the following two steps: i) If  $\abf_j$ in \eqref{eqn:ajFromSubAQoS}  under $\tilde{\lambda}_j=0$ satisfies the constraint in \eqref{eqn:ADMMpSubA-QoS}, then   it is the optimal solution, and $\tilde{\lambda}^o_{j}=0$; ii) otherwise, $\tilde{\lambda}^o_{j}$ is  such that \eqref{eqn:ADMMpSubA-QoS} holds with equality. We can use the bi-section search for $\tilde{\lambda}^o_{j}$   such that $\frac{1}{p_i} \| \Gbf_{i} \abf_{i} \|^{2} = v$.

\subsubsection{Algorithm  Convergence}
In summary, our proposed fast algorithm for computing  $\abf_i$ in $\Pc_2$ is a two-layer iterative algorithm. It consists of the outer-layer SCA iterations and the inner-layer ADMM iterations for solving each SCA subproblem $\Pc_{2\SCA}(\ubf)$ (see  Fig.~\ref{alg2_CPU} for the flow diagram).   The  updates  in \eqref{eqn:ADMM-update-1} -- \eqref{eqn:ADMM-update-4} at each ADMM iteration are all computed in closed-form  as  in \eqref{eqn:ADMM-dj,ikQoS},  \eqref{eqn:ADMM-vExpressQoS} and \eqref{eqn: t_opt}, or semi-closed-form as in  \eqref{eqn:ajFromSubAQoS}, respectively.

As mentioned earlier, since  $\Pc_{2\SCA}(\ubf)$ is convex, our inner-layer ADMM-based algorithm    in \eqref{eqn:ADMM-update-1} -- \eqref{eqn:ADMM-update-4} is guaranteed to converge to the optimal solution of   $\Pc_{2\SCA}(\ubf)$ \cite{Boyd&etal:2011ADMM}.
Following this, the outer-layer SCA iteration is guaranteed to converge to the stationary point of $\Pc_2$ \cite{Marks&Wright:OR78}.

%%%%%%%%%%%%%%%
\subsection{Semi-Distributed Computing Approach for  $\wbf_i$} \label{subsec:semi-distr}
Note that   $\abf_i$'s are jointly optimized in $\Pc_2$ for the coordinating BSs, which requires centralized processing.  Typically, a centralized method requires the global CSI, \ie all $M\times 1$ user channel vectors $\{\hbf_{i,jk}\}$ from all $J$ BSs, and the data exchange overhead in terms of the number of complex scalars is $MKJ^2$, which is substantial especially for massive MIMO with $M\gg 1$. For the network architecture such as Cloud-Radio Access Network (C-RAN), such  data exchange between the centralized processing unit (CPU) and the BSs  requires high-bandwidth and low-latency fronthaul communication and imposes challenges for real-time coordination. 

We show that our proposed  algorithm does not require the global CSI exchange. Below, we present  a semi-distributed  computing approach to carry out the algorithm between the CPU and the BSs efficiently using the local CSI at each BS: 
\begin{mylist}
\item {\bf Computation at the CPU:} The CPU\ uses the proposed ADMM-based algorithm to compute     $\{\abf_i\}$ centrally using the  iterative updates  in \eqref{eqn:ADMM-update-1} -- \eqref{eqn:ADMM-update-4}. Examining the expressions for these updates, \ie $\dbf_{ik}^{o}$ for  $\Pc_{\dbf_{ik}}(\ubf)$ in \eqref{eqn:ADMM-dj,ikQoS}, $v^o$ in  \eqref{eqn:ADMM-vExpressQoS}, $t^o$ in \eqref{eqn: t_opt}, and $\abf_i$ in  \eqref{eqn:ajFromSubAQoS}, we notice that the CPU only needs  $\Gbf_i^H\Gbf_i$ and $\{\fbf_{i,jk}$, $k\in \Kc$, $j\in \Jc\}$ from each BS $i$ to compute the updates. They are  $K\times K$ matrix and $K\times 1$ vectors. These quantities can be computed locally at each BS $i$ based on the local CSI $\{\hbf_{i,jk},  k\in \Kc, j\in \Jc\}$ (see below). 
Therefore, instead of  obtaining  the global CSI $\{\hbf_{i,jk}\}$ from all BSs,  the CPU only obtain these necessary quantities from  each BS $i$ and then compute $\abf_i$'s using the proposed  algorithm. 

\item {\bf Computation at each BS:}   At BS $i$, once  $\Rbf_i(\lambdabf, \mu_i)$ is obtained locally  as discussed in Section~\ref{subsubsec:R}, the BS  compute  $\Gbf_i=\Rbf^{-1}_{i}(\lambdabf, \mu_i)\Hbf_{i}$, and then $\Gbf_i^H\Gbf$ and  $\fbf_{i,jk}=\Gbf_i^H\hbf_{i,jk}$,   based on the local CSI \{$\hbf_{i,jk}$, $k\in \Kc$, $j\in \Jc\}$.
Then, each BS sends $\Gbf_i^H\Gbf_i$  and $\{\fbf_{i,jk},  k\in \Kc,j\in\Jc\}$ to the CPU. Once the CPU obtains $\abf_i$'s, it sends  $\abf_i$ to each BS $i$. Then, BS $i$ generates $\wbf_i$ using   \eqref{eqn:thm1_w_opt}.
\end{mylist}
% %%%%%%%%%%%%%%%%%%%%%%%%%%%%%%%%%%%%%%%%%
\begin{algorithm}[t]
\begin{algorithmic}%[1]
\caption {The Fast Algorithm with Semi-Distributed Computing  for Coordinated Multicast Beamforming Problem $\Pc_o$}   \label{algADMMSCA}
  \State \hspace*{-1em}\textbf{I) At each BS $i$:}
 \State Compute  $\Rbf_i\left( \lambdabf, \mu_i\right)$ by Algorithm~\ref{alg:LambdaObtainQoS}.
 \State Compute $\Gbf_i^H\Gbf_i$ and $\fbf_{i,jk}$, $\forall k\in\Kc, j\in\Jc$.
 \State Send  $\Gbf_i^H\Gbf_i$ and $\{\fbf_{i,jk}$, $\forall k\in\Kc, j\in\Jc\}$ to the CPU.
\vspace*{.5em}
  \State \hspace*{-1em}\textbf{II) At the CPU:}
{ \State \textbf{Initialization}: Generate initial point $\ubf$. Set $\rho$.
\Repeat  ~// Outer-layer 
 \State \textbf{Initialization}: Set $\abf^{(0)}\!=\ubf$. Set $\qbf^{(0)}\!, z^{(0)}\!,t^{(0)}$\!. Set $l=0$.
\Repeat  ~// Inner-layer for solving $\Pc_{2\SCA}(\ubf)$
\State \hspace*{-1em} 1) Update  $\dbf_{ik}^{(l+1)}$ via \eqref{eqn:ADMM-dj,ikQoS}, $\forall k\in\Kc, i\in\Jc$.
\State \hspace*{-1em} 2) Update  $v^{(l+1)}$ via  \eqref{eqn:ADMM-vExpressQoS}.
\State \hspace*{-1em} 3) Update   $\abf_i^{(l+1)}$  via  \eqref{eqn:ajFromSubAQoS}, $\forall i \in \Jc$.
\State \hspace*{-1em} 4) Update $t^{(l+1)}$ via
 \eqref{eqn: t_opt}. 
 \State \hspace*{-1em} 5) Update  $\qbf^{(l+1)}$ via \eqref{eqn:ADMM-update-3} and $z^{(l+1)}$ via \eqref{eqn:ADMM-update-4}.
\State \hspace*{-1em} 6) Set $l\leftarrow l+1$.  
\Until convergence.
\State Set $\ubf = \abf^{(l+1)}$. 
\Until convergence.
 \State  Send $\abf_i^{(l)}$ to BS $i$, for $i\in \Jc$.
}
\vspace*{.5em}
  \State \hspace*{-1em}\textbf{III) At each BS $i$:}
  \State Compute $\wbf_i$ via  \eqref{eqn:thm1_w_opt}. 
\end{algorithmic}
\end{algorithm}

\begin{figure}[t]
\vspace*{-1em}
\centering
\includegraphics[width=.85\columnwidth]{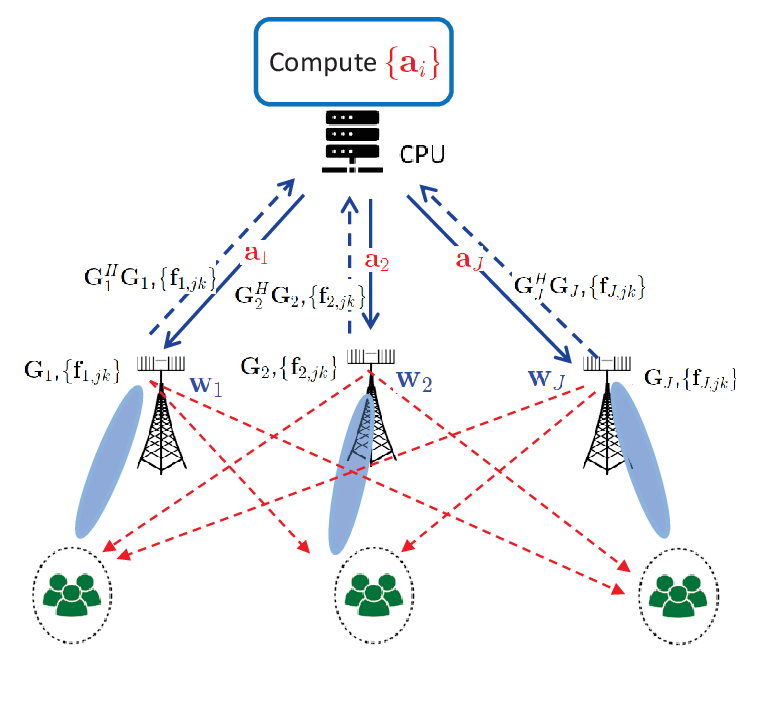}\vspace*{-1.5em}
\caption{The illustration of the semi-distributed computing approach of Algorithm~\ref{algADMMSCA} for coordinated multicast beamforming among BSs.}\vspace*{-.5em}
\label{alg2_diagram}
\end{figure}

We summarize our proposed fast algorithm for coordinated multicast beamforming in Algorithm~\ref{algADMMSCA}, and  the semi-distributed computing approach  in  Algorithm~\ref{algADMMSCA} is shown in Fig.~\ref{alg2_diagram} . This approach  explores the essential information required from each BS and integrates  the computational capability of both   the BSs and the CPU. As a result, it significantly reduces the amount of information exchanged through fronthaul communication to generate the beamformers $\wbf_i$'s.\footnote{Note that we can implement a fully distributed algorithm by moving the computation of $\dbf_{ik}$'s and $\abf_i$'s to each BS without the need for BS sending $\Gbf_i^H\Gbf_i$ and $\{\fbf_{i,jk}\}$. This would require some limited information exchange between each BS and the CPU for the updates. However, since these quantities need to be updated iteratively, this approach could cause fronthaul delay, which is undesirable. Therefore we prefer conducting the main iterative algorithm  at the CPU using a semi-distributed approach.  } 
\begin{figure}[t]
%\vspace*{-.5em}
\centering
\includegraphics[width=.65\columnwidth]{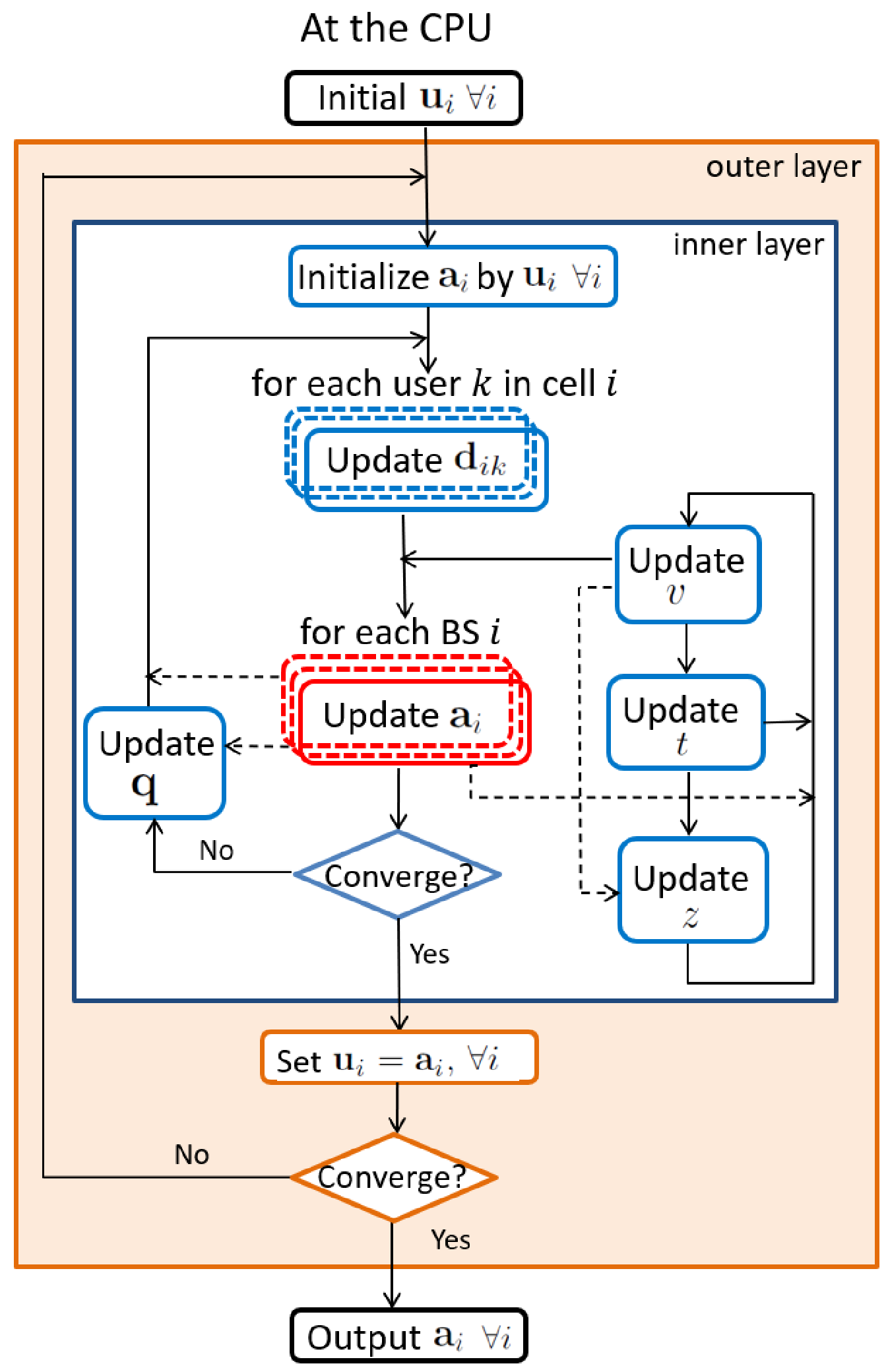}
\caption{The flow diagram of computing $\{\abf_i\}$  at the CPU in Stage II of Algorithm~\ref{algADMMSCA}.}\vspace*{-.5em}
\label{alg2_CPU}
\end{figure}

Furthermore, for the main computation at the CPU in Algorithm~\ref{algADMMSCA}, a flow diagram of the two-layered iterative algorithm is shown in Fig.~\ref{alg2_CPU}.  In particular, we point out that the updates of  $\dbf_{ik}$'s can   be computed in parallel for each   $k$ and $i$, since each is the solution of a separate subproblem $\Pc_{\dbf_{ik}}(\ubf)$. The same applies to  $\abf_i$'s,  which can all be computed in parallel using  \eqref{eqn:ajFromSubAQoS}. This feature in our proposed algorithm provides a further computational advantage for  practical implementation, where the computational time for these two main updates $\{\dbf_{ik}\},\{\abf_i\}$  will not increase with $J$ if parallel computing is employed.  

In summary,  Algorithm~\ref{algADMMSCA} is efficient in both  computation and communication. Its computational complexity and fronthaul communication overhead are analyzed below.

\subsubsection{Computational Complexity Analysis} \label{subsec:complexity}
The main computation in
Algorithm~\ref{algADMMSCA} is  the two-layer iterative algorithm carried out at the CPU.  Each inner-layer iteration involves five updates: (1) Updating each $\dbf_{ik}^{(l+1)}$ using \eqref{eqn:ADMM-dj,ikQoS} requires  $2JK+\text{const}\cdot J$ flops\footnote{In \eqref{eqn:ADMM-dj,ikQoS}, we only need to compute $e_{1,j,ik}^{(l+1)}$, $j\in\Jc$. The rest of values are fixed and can be computed at the beginning of each SCA iteration.}. 
Note that all  $\dbf_{ik}^{(l+1)}$'s can be computed in parallel, where the time complexity can be similar to that of computing each  $\dbf_{ik}^{(l+1)}$. (2) Updating $v^{(l+1)}$ in (35) requires $J(K^2+K)+J$ flops. The computation mainly is from calculating  $\abf_i^{(l+1)H}(\Gbf_i^H\Gbf_i)\abf_i^{(l+1)}$, for each $i\in \Jc$, where   $\Gbf_i^H\Gbf_i$ is  provided at the CPU.
 Thus, the leading complexity in this update is $JK^2$ flops. (3) Updating $\abf_i^{(l+1)}$ in (41) depends on $\tilde{\lambda}_i$ value. If $\tilde{\lambda}_i=0$, then the leading complexity is $JK^2$ flops. Note that the  matrix inversion in this case  involves fixed values and only needs to be performed once   at the beginning of the algorithm.  If $\tilde{\lambda}_i>0$, we need to perform matrix inversion with complexity $I_{a}(O(K^3)+K^2)+JK(K+1)$ flops, where $I_a$ is the number of bi-section searches required. Thus, the leading complexity for computing each $\abf_i^{(l+1)}$ is either  $JK^2$ flops or $O(K^3)$ in the worst case. 
Again, note that all   $\abf_i^{(l+1)}$'s can be computed in parallel with the time complexity being similar to that of computing each $\abf_i^{(l+1)}$. (4) Updating  $t^{(l+1)}$ and  $\qbf^{(l+1)}$ are straightforward and requires about  $J^2K$ flops.  

Thus, for each inner-layer iteration, the main computation occurs at updating $\abf_i$'s in (41). The overall leading time complexity  per iteration, assuming
parallel computing can be implemented, is similar to that of the equivalent computational complexity of $\text{const}\cdot [JK^2+JK]$ flops in the best case or $O(K^3)+ \text{const}\cdot JK^2$ flops in the worst case.

From the above analysis, the computational complexity of the main  algorithm at the CPU is independent of the number of BS antennas $M$ and grows  linearly with $J$  coordinating BSs. This is attractive for massive MIMO systems with a large value of $M$, and  further increasing $M$    will not affect the algorithm complexity. At the same time, the algorithm allows more BSs to participate in coordination with only a mild growth of complexity.

\subsubsection{Fronthaul Communication Overhead Analysis}\label{subsec:overhead}
In Algorithm~\ref{algADMMSCA}, the information exchange between the BSs and the CPU occurs in  three stages: 
\begin{itemize}
\item[i)] Computing $\Rbf_i(\lambdabf, \mu_i)$ by Algorithm~\ref{alg:LambdaObtainQoS} at BS $i$; 
\item[ii)]  BS $i$ sends   $\Gbf_i^H\Gbf_i$ and $\{\fbf_{i,jk}$,  $k\in\Kc, j\in\Jc\}$ to the CPU;
\item[iii)] The CPU sends $\abf_i^{(l)}$ to each BS $i$.
\end{itemize}

For i), Algorithm~\ref{alg:LambdaObtainQoS} needs to exchange $K\times 1$ vector $\lambdabf_i^{(m)}$'s among $J$ BSs in each iteration. As discussed in Section~\ref{subsubsec:R},  this requires exchanging $JK$ real values per iteration. Our simulation study shows the number of iterations is typically about $5\sim 15$  for $M$ ranging from 100 to $200$. For ii) and iii), note that $\Gbf_i^H\Gbf_i$ is a $K\times K$ matrix, and both $\fbf_{i,jk}$ and $\abf_i$ are  $K\times 1$ vectors. The total information exchange between the CPU and all BSs in terms of the number of complex scalars is $K^2J(J+1)+KJ$, which does not depend on the number of BS antennas $M$. 

 Thus, the entire information exchange  required via fronthaul by Algorithm~\ref{algADMMSCA}   in terms of complex scalars is  $K^2J(J+1)+ \text{const}\cdot JK$, which is independent of $M$. This is particularly beneficial for massive MIMO, as the communication overhead    is significantly lower than $MKJ^2$  for the conventional centralized processing, and
the  communication saving  becomes more significant as  $M$ becomes larger.  Since the total information exchange does not grow with $M$, increasing  the number of antennas at the BSs will not  impact  the fronthaul  requirement in terms of both capacity and delay. Overall, the significant reduction of communication overhead  further allows more BSs to participate in coordination.

From the analysis in Sections~\ref{subsec:complexity} and \ref{subsec:overhead}, it is apparent that the proposed algorithm is highly efficient in both computation and communication: both computational complexity and amount of information sharing will remain unchanged when the number of BS antennas further increases, as expected in the future systems with ultra-massive MIMO. These efficiencies encourage more BSs to participate in coordination to further reduce interference and improve the overall system performance.

\subsubsection{Initialization}:
For the initial point $\ubf$ in the SCA method  to  solve
$\Pc_{2\SCA}(\ubf)$ iteratively, different conventional initialization methods can be used. In particular, since the converted problem   $\Pc_2$ has a much smaller problem size with the original $\Pc_1$, we can apply the conventional SDR  along with the Gaussian randomization method  to find a feasible point for $\Pc_2$ to be used as the initial point.
%\footnote{We can limit the number of trials in the Gaussian randomization procedure, as we only need a feasible solution.} 
We note that following our method above, the  CPU has all the information obtained from BSs to compute the initial point.

\section{Other Coordination  Conditions or Scenarios} \label{sec:others}
\subsection{Coordinated Multicasting under Imperfect CSI}
So far, we have assumed perfect CSI in deriving the optimal beamforming structure and proposing the fast semi-distributed algorithm to generate $\wbf_i$ at each BS $i$. In practice, each BS only has the estimated local CSI available. Below, we show how our results and proposed approach can be extended to incorporate the imperfect CSI. 

Consider each channel $\hbf_{i,jk}$ follows a general Rayleigh fading distribution as   $\hbf_{i,jk}\sim \Cc\Nc({\bf 0}, \Cbf_{i,jk})$, where $\Cbf_{i,jk}$ is the channel covariance matrix. Let  $\hat{\hbf}_{i,jk}$  be the MMSE estimate of $\hbf_{i,jk}$. The estimation error $\tilde{\hbf}_{i,jk}=\hbf_{i,jk}-\hat{\hbf}_{i,jk}$ is independent to $\hat{\hbf}_{i,jk}$ and has the following distribution $\tilde{\hbf}_{i,jk}\sim \Cc\Nc({\bf 0}, \Ebf_{i,jk})$, where $\Ebf_{i,jk}$ is the covariance matrix for the estimation error. For  the downlink  massive MIMO,  using the capacity lower bound, an achievable rate at user $k$ in cell $i$   is given by $\log(1+\SINR_{ik}^\text{\tiny eff})$, where $\SINR_{ik}^\text{\tiny eff}$ is the effective SINR given by \cite{Bjornsonetal:Foundation17}
\begin{align}
\hspace*{-.5em}\SINR_{ik}^\text{\tiny eff} &= 
\frac{|\mathrm{E}(\wbf_i^H\hbf_{i,ik})|^{2}}
{ \sum^{J}_{j=1}\mathrm{E}(|{\wbf}_{j}^H\hbf_{j,ik}|^2)-|\mathrm{E}(\wbf_i^H\hbf_{i,ik})|^2+\sigma^2}
\end{align}
 Consider the BS evaluates the above effective SINR  given all the MMSE estimates $\{\hat{\hbf}_{i,jk}\}$, which we refer to as the instantaneous effective SINR at each user that is perceived by the BSs. It is given by
\begin{align}\label{SINR_est}
\hspace*{-.5em}{\SINR}^\text{est}_{ik} &=\frac{|{\wbf}_{i}^H\hat{\hbf}_{i,ik}|^{2}}{ \sum^{J}_{{j=1,j\neq i}}\!|{\wbf}_{j}^H\hat{\hbf}_{j,ik}|^2+\sum^{J}_{j=1}\wbf_j^H\Ebf_{j,ik}\wbf_j+\sigma^2}.
\end{align}       
Then, the original coordinated multicast beamforming problem $\Pc_o$   is modified to the following
 \begin{align} %\label{eqn:QoSproblem0}
\Pc_{o}^\text{est}:\ \min_{\Wbf} \max_{i}  \ & \frac{1}{p_i}\|\wbf_{i}\|^{2}\nn \\
\st\  &{\SINR}^\text{est}_{ik} \geqslant\gamma_{ik}, \ k \in \Kc, i \in \Jc \nn 
\end{align}
where  SINR in the constraints  is replaced with the perceived instantaneous effective SINR in \eqref{SINR_est} for the BSs to jointly optimize $\wbf_i$'s.

Compared with  \eqref{eqn:SINRexpression}, the SINR expression in \eqref{SINR_est} has an additional second term in the denominator (also, each channel is replaced by its estimate), which reflects the uncertainty due to the estimation errors of channels from all BSs to a user. Nonetheless, the structure in SINR expression w.r.t. $\{\wbf_i\}$ still maintains  the same, and all our previous derivations in Section~\ref{OptimalStructureQoS}  leading to Theorem~\ref{theorem-1QoS} can be straightforwardly adapted to the new SINR expression. Following this, the optimal coordinated multicast beamforming solution for $\Pc_{o}^\text{est}$ is given by the following corollary.
\begin{corollary}
Based on  the MMSE channel estimates at all the BSs, the optimal  solution to the QoS problem $\Pc_{o}^\text{est}$ for  coordinated multicast beamforming is given by
 \begin{align}\label{eqn:w_opt_est}
\wbf_{i} 
&=\hat{\Rbf}^{-1}_{i}( \lambdabf, \mu_i)\hat{\Hbf}_{i}\abf_{i},\ i \in \Jc
\end{align}
where 
\begin{align}\label{eqn:R_est}
\hspace*{-.5em}\hat{\Rbf}_i\left( \lambdabf, \mu_i\right)\triangleq \frac{\mu_{i}}{p_i}\Ibf+\sum^{J}_{j= 1}\sum^{K}_{k=1}\lambda_{jk}\gamma_{jk}\big(\hat{\hbf}_{i,jk}\hat{\hbf}^{H}_{i,jk}+\Ebf_{i,jk}\big),
\end{align}
and each weight in $\abf_i$ is $a_{ik} = \lambda_{ik}(1+\gamma_{ik})(\hat{\hbf}^H_{i,ik}\wbf_{i})$.
\end{corollary}
Note that the beamforming structure in \eqref{eqn:w_opt_est} is the same as that in \eqref{eqn:thm1_w_opt} of the perfect CSI case, except that compared with $\Rbf_i(\lambdabf,\mu_i)$ in \eqref{eqn:covarianceMatrixRwithMuQoS}, the summation term in $\hat{\Rbf}_i\left( \lambdabf, \mu_i\right)$ for each $\hat{\hbf}_{i,jk}$ contains an additional  covariance term  $\Ebf_{i,jk}$  that captures the estimation error. Thus, the discussions in Remarks~\ref{remark1}-\ref{remark3} on the optimal beamforming structure also apply here to  the imperfect CSI case. In particular, since each BS $i$ has the local channel estimates and the corresponding estimation error covariance matrices, $\{\hat{\hbf}_{i,jk}, \Ebf_{i,jk}, \forall k, j\}$, the optimal beamformer $\wbf_i$ in \eqref{eqn:w_opt_est} is still a \emph{distributed} beamformer that can be computed locally at BS $i$. 

Furthermore, our proposed Algorithms~\ref{alg:LambdaObtainQoS} and \ref{algADMMSCA}, including the fast algorithm for weights $\{\abf_i\}$ and the semi-distributed computing approach to generate $\wbf_i$ at each BS $i$, can be directly extended to the\ estimated CSI case. Specifically, in these algorithms, all the computations using channel $\hbf_{i,jk}$ can be replaced with estimate $\hat{\hbf}_{i,jk}$, and  $\Rbf_i\left( \lambdabf, \mu_i\right)$ is replaced with $\hat{\Rbf}_i\left( \lambdabf, \mu_i\right)$. The details of the algorithms under the estimated CSI are omitted to avoid repetition.

\subsection{Extension to Other Coordination Scenarios}\label{subsec:others}
\subsubsection{Multiple Groups per Cell} Our system model assumes one group per cell to keep the exposition simple. The results can be extended directly to the general case that includes  $G_i$ multiple groups in each cell $i$,  with $K_g$ users in group $g$. In this case, the total transmit power at BS $i$ is given by $\sum_{g=1}^{G_i} \|\wbf_{ig}\|^2$, where $\wbf_{ig}$ is the multicast beamformer for group $g$ in cell $i$.  It is essentially an instance of the single-cell multi-group scenario considered in \cite{Dong&Wang:SP20} if only BS $i$ is considered. For coordination among BSs, the transmit power constraint in \eqref{eqn:Pwr_constr} is changed to $\frac{1}{p_i} \sum_{g=1}^{G_i} \!\|\wbf_{ig}\|^2 -t\leqslant0$ for each BS $i$. In this case, the SINR expression in   \eqref{eqn:SINRexpression} also contains intra-cell inter-group interference at the denominator. All the derivations in Section~\ref{OptimalStructureQoS}  leading to Theorem~\ref{theorem-1QoS} can still be straightforwardly adapted to this SINR expression, and the optimal multicast beamformer
for group $g$ in cell $i$ is  
\begin{align}
\wbf_{ig} = \Rbf^{-1}_{i}( \lambdabf, \mu_i)\Hbf_{ig}\abf_{ig} 
\end{align}
where $\Hbf_{ig}$ is the channel matrix between BS $i$ and  user group $g$ in cell $i$, and  $\abf_{ig}$ is the weight vector for this group.
Also, $\lambdabf$ now has the dimension of the total number of users in the system,   $\sum_{i=1}^J\sum_{g=1}^{G_i}K_g$, with element $\lambda_{igk}$ associated with each SINR constraint for user $k$ in group $g$ in cell $i$. Again, our proposed Algorithms~\ref{alg:LambdaObtainQoS} and \ref{algADMMSCA} can be straightforwardly extended to this case to compute $\{\abf_{ig}\}$ at the CPU, and each BS $i$ distributively generates the multicast beamformers $\{\wbf_{i1},\ldots,\wbf_{iG_i}\}$ for $G_i$ groups based on the local CSI.
%%%%%%%%%%%%%%%%%%%%%
\subsubsection{Coordination among BS Clusters}
So far, we have assumed that each BS serves its own users and coordinates with other BSs. To further improve the performance, BS clustering may be considered, 
where a subset of BSs  fully cooperate to jointly serve their users. For full cooperation, data sharing among BSs in a cluster is required for the BSs to form joint multicast beamforming to serve their users, and coordinated beamforming  among BS clusters is performed for managing inter-cluster interference. 
When the BS clusters are disjoint, \ie each BS only participate in one cluster, each BS cluster can be effectively viewed as a ``super" BS with distributed antennas as in our system model. It is easy to see that our results and proposed algorithms can be directly applied to this case for coordinated multicast beamforming among BS clusters, where each BS cluster can generate its respective beamformers distributively without global CSI sharing among different BS clusters.\footnote{Within a BS cluster, CSI sharing among the BSs in the cluster may be required for joint beamforming to maximize the full cooperation gain.} 
%%%%%%%%%%%%%%%%%%%%%%%%%%%%%%%%%%%%%%%%%%%%%%%%%%%%%%%%%
\section{Simulation Results} \label{sec:sim}
%%%%%%%%%%%%%%%%%%%%%%%%%%%%%%%%%%%%%%%%%%%%%%%%%%%%%%%%%
We consider a coordinated multi-cell  multicast beamforming scenario with $J=3$ BSs and one group per cell. Each cell has a unit cell radius, and users in the cell are randomly located with a uniform distribution. All user channels are generated independently, each follows a complex Gaussian distribution $\hbf_{i,jk} \sim\Cc\Nc(\mathbf{0}, \beta_{i,jk} \bf{I})$, $\forall k,i,j$. The channel variance    $\beta_{i,jk}$ is modeled by the path loss model:  $\beta_{i,jk} = \xi_{0}d^{-\kappa}_{i,jk}$, where $d_{i,jk}$ is the distance between BS $i$ and  user $k$ in cell $j$,  the pathloss exponent is $\kappa=3.5$, and $\xi_{0}$ is the path loss constant. The value of $\xi_0$ is determined by setting the nominal average received SNR  under a unit transmit power at the cell boundary to be $-5$ dB, \ie $\frac{\xi_0}{\sigma^2}  = -5$ {dB}. We set  the power budget target of each BS as $p_i=10$~dBW, $i\in\Jc$. The performance results are averaged over $100$ channel realizations and $10$ realizations of user locations.

%%%%%%%%%%%%%%%%%%%%%%%%%%%%%%%%%%%%%%%%%%%%%%%%%%%%%%%%%
\subsection{Convergence Behavior}\label{sec:ConvergenceAna}
%%%%%%%%%%%%%%%%%%%%%%%%%%%%%%%%%%%%%%%%%%%%%%%%%%%%%%%%%
%\subsubsection{For QoS Problem $\Pc_o$}
We first show the convergence behaviour of our proposed fast algorithm for solving $\Pc_{o}$.  It is based on the optimal beamforming structure in \eqref{eqn:thm1_w_opt} and solving $\Pc_2$ via  Algorithm~\ref{algADMMSCA}. The main algorithm carried out at the CCU 
consists of the outer-layer SCA iteration over $\ubf$, and the inner-layer ADMM-based iteratively updates for solving $\Pc_{2\SCA}(\ubf)$ in each SCA iteration. We set the penalty parameter $\rho = 0.01$.\footnote{We have conducted extensive experiments to study the effect of different values of  $\rho$ on the performance and selected this value, which provides the best trade-off of performance and convergence speed.}   

\begin{figure}[t]
        \centering
        \includegraphics[width=.8\columnwidth]{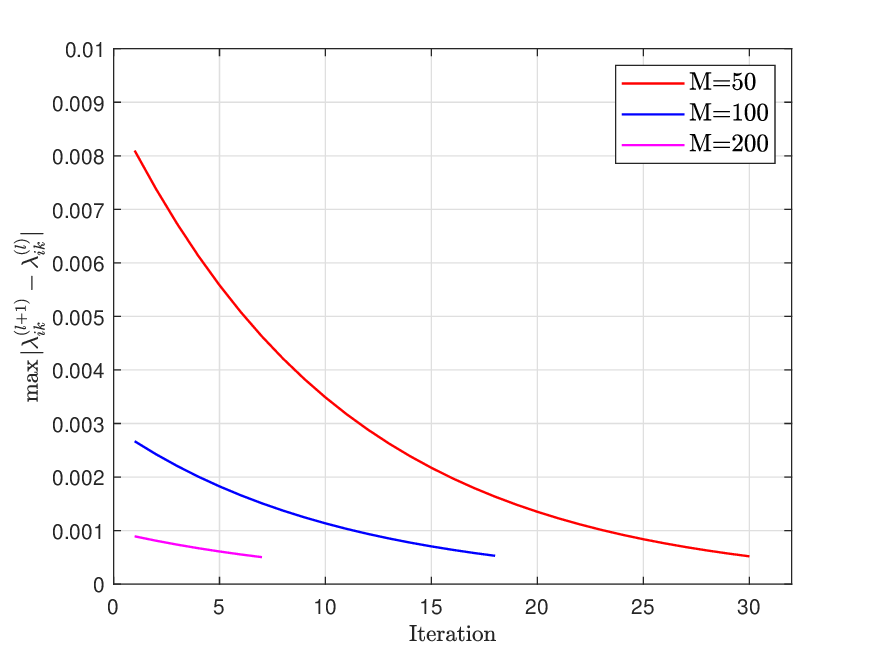}
        \centering
        \caption{The convergence of $\widetilde{{\lambdabf}}$ by using Algorithm~\ref{alg:LambdaObtainQoS} ($J=3$, $K=5$).}
        \label{fig:lambdaConvergence}    
%\end{figure}
%\begin{figure}[H]
        \centering
        \includegraphics[width=.8\columnwidth]{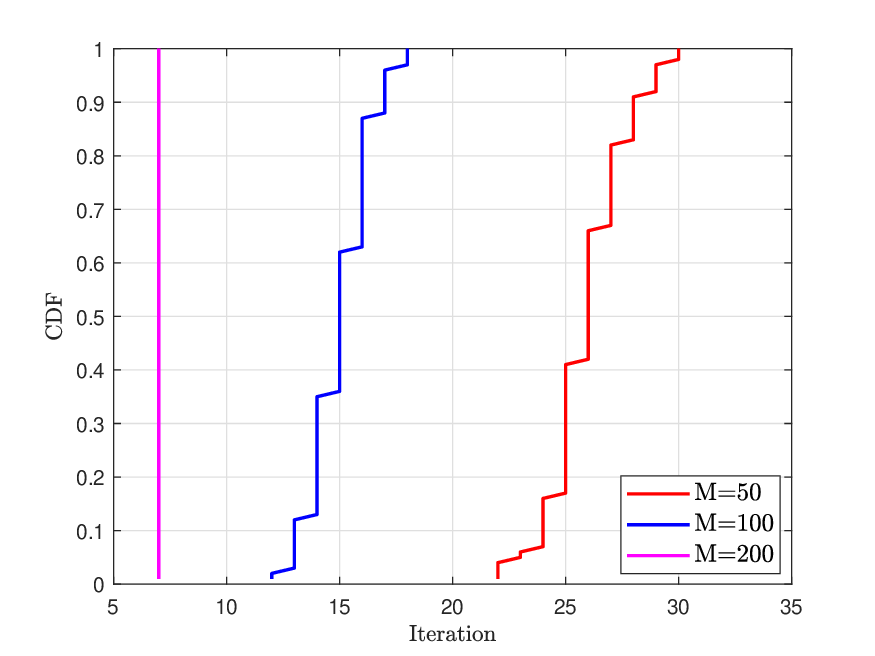}
        \centering
        \caption{The empirical CDF of the iterations need for $\lambdabf$ convergence under different $M$ ($J=3$, $K=5$).}
        \label{fig:lambdaCDF} \vspace*{-1em}
\end{figure}

We first study the convergence behaviour of Algorithm~\ref{alg:LambdaObtainQoS} for computing $\lambda_{ik}$'s.  Fig.~\ref{fig:lambdaConvergence} shows the convergence behaviour in terms of the maximum difference $\max_{i,k}|\lambda_{ik}^{(l+1)}-\lambda_{ik}^{(l)}|$ over iterations $l$ for $M=50, 100,200$. We set $K=5$. 
We see that the maximum difference of $\lambda_{ik}$'s drops  below $0.5\times 10^{-3}$ in less than $30$ iterations for $M=50$. As $M$ increases, the convergence rate becomes faster, and only less than $10$ iterations are needed for $M=200$. To show the statistical information on the convergence rate, in Fig.~\ref{fig:lambdaCDF}, we plot the empirical cumulative density function (CDF) of the number of iterations required for the maximum difference  $\max_{i,k}|\lambda_{ik}^{(l+1)}-\lambda_{ik}^{(l)}| \leqslant0.5\times 10^{-3}$ generated over $100$ channel realizations. We see that Algorithm~\ref{alg:LambdaObtainQoS} typically converges within $30$ iterations to $5$ iterations for $M$ ranges from $50$ to $200$, which is consistent with Fig.~\ref{fig:lambdaConvergence}. The convergence tends to be come faster as $M$ increases. This could be that the expression at the right hand side of \eqref{lambda:fixed-point} converges to an asymptotic value, which expedites the fixed-point convergence. 

\begin{figure}[t]
\centering
\includegraphics[width=.8\columnwidth]{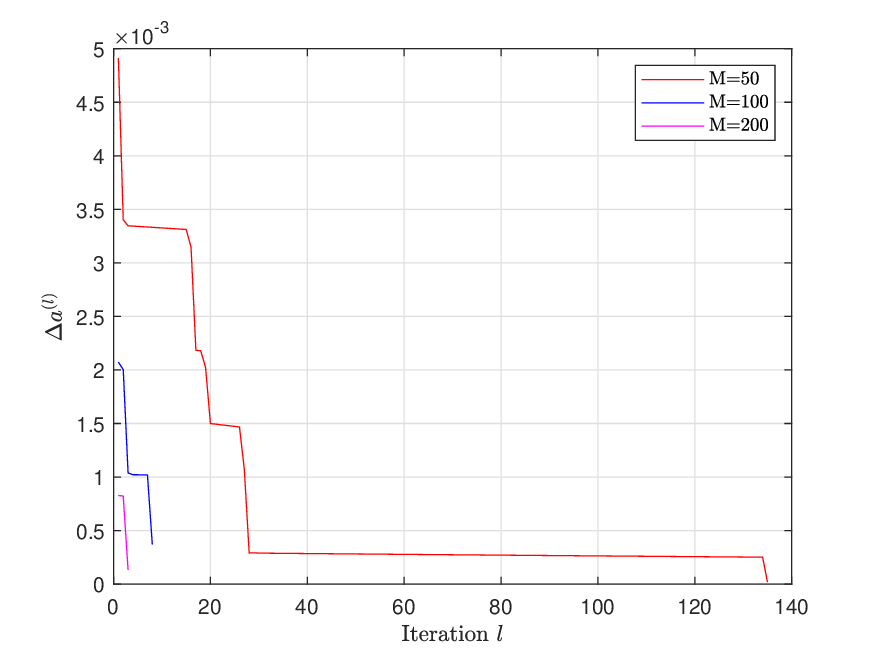}
\centering
\caption{Convergence behaviour of the inner-layer algorithm at the CCU in Algorithm~\ref{algADMMSCA}: the maximum relative difference $\Delta a^{(l)}$ over iteration $l$ (In the first outer-layer SCA iteration. $K = 5$).}
\label{fig:aInner_ADMM}
%\end{figure}
%\begin{figure}[t]
        \centering
        \includegraphics[width=.8\columnwidth]{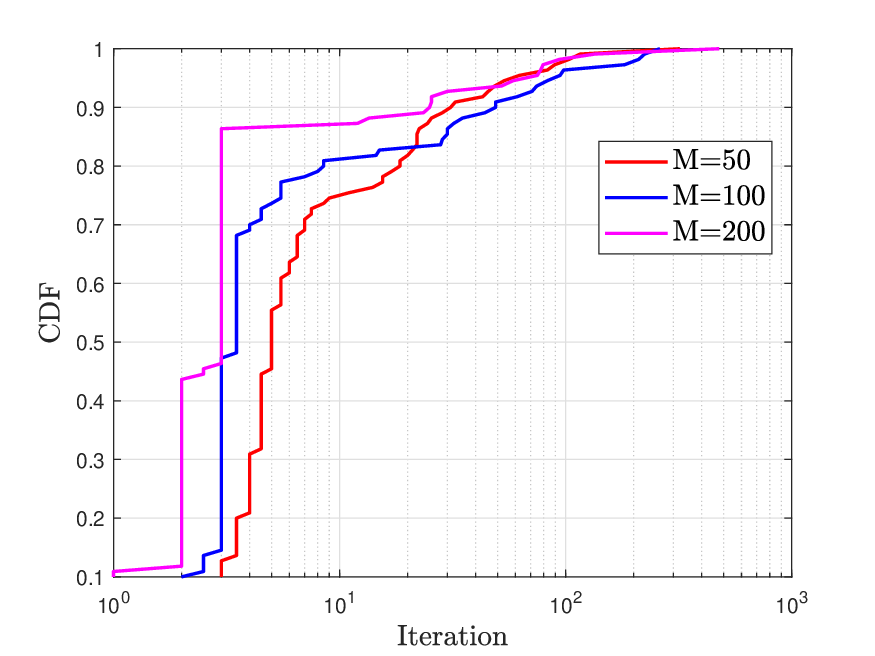}
        \centering
        \caption{The empirical CDF  of the number of inner-layer iterations required for $\Delta a^{(l)} \leqslant 10^{-3}$ ($K = 5$).}
        \label{fig:admmCDF} 
%\end{figure}
%\begin{figure}[t]  
\centering
\includegraphics[width=.8\columnwidth]{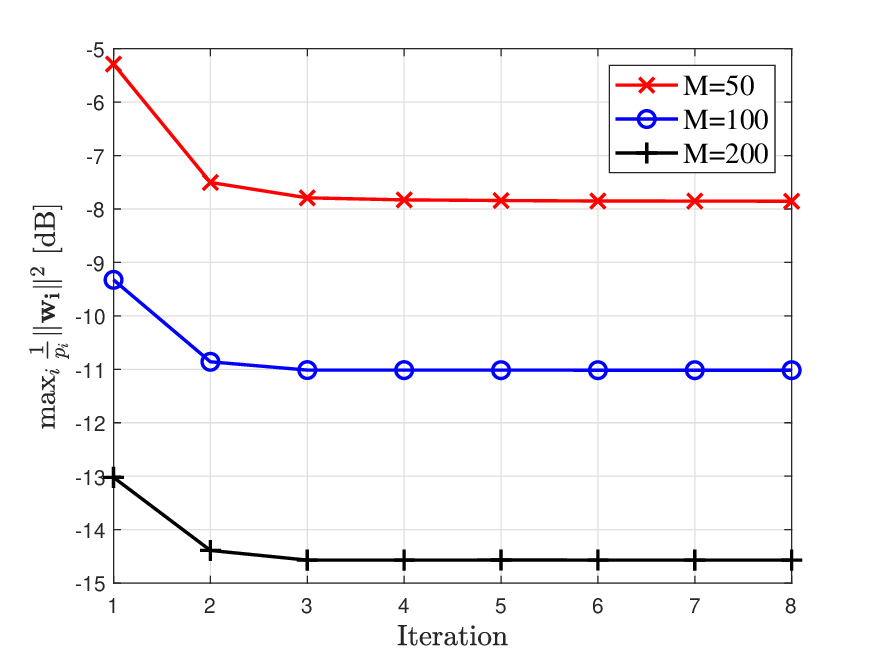}
\centering
\caption{ Convergence behaviour of the outer-layer algorithm at the CCU in Algorithm~\ref{algADMMSCA}: the objective $\max_i \frac{1}{p_i}\|\wbf_i\|^2$ over the  SCA iterations ($ K = 5$).}
\label{fig:tOuter_ADMM}\vspace*{-1em}
\end{figure}

We now study the convergence behavior of the inner-layer iterations in Algorithm~\ref{algADMMSCA}. We define the maximum relative difference of $\abf^{(l)}$ between two consecutive iterations as $\Delta a^{(l)} \triangleq \max_{i\in \Jc}\frac{\|\abf^{(l+1)}_{i}-\abf^{(l)}_{i}\|}{\|\abf^{(l)}_{i}\|}$. Fig.~\ref{fig:aInner_ADMM} left
shows the convergence behaviour of  $\Delta a^{(l)}$ over iterations in the first outer-layer SCA iteration, for $M=50, 100, 200$. We set $K=5$ users per group.  We see that the value of $\Delta a^{(l)}$ drops fast, especially when $M$ becomes large. Typically,  it drops below $1\times 10^{-3}$ in less than $10$ iterations for $M\geqslant 100$ and  $\sim20$ iterations for $M=50$. For further reducing the value of 
$\Delta a^{(l)}$, more iterations may be required  for $M=50$, while much fewer iterations are used for $M\geqslant 100$.\footnote{We observe that these convergence curves decrease slowly then have a big drop. Our explanation for this  is that the algorithm searches for different directions in  $\abf_i^{(l)}$  to reduce the objective value, and the sudden drop indicates an effective direction is found, which leads to a large reduction in the objective value.} Note that  as the outer-layer SCA iteration increases, $\ubf$ converges to $\abf$, and as a result, the inner-layer  convergence becomes even faster for computing the solution to $\Pc_{2\SCA}(\ubf)$.
Fig.~\ref{fig:aInner_ADMM} shows the convergence behavior using some random channel realizations. To see the statistical convergence behavior, we plot the empirical CDF of the number of iterations needed for  $\Delta a^{(l)}$ to drop below  $1\times 10^{-3}$ in the first outer-layer SCA iteration, as shown in Fig.~\ref{fig:admmCDF}. We see that the average convergence rate becomes slightly faster as $M$ increases. Over $90\%$ channel realizations can converge less than $10$ iterations, and over $95\%$ channel realizations can converge less than $100$ iterations. 

In Fig.~\ref{fig:tOuter_ADMM}, we show the trajectory of the objective value of $\Pc_2$ over the outer-layer SCA iterations computed at the CPU in Algorithm~\ref{algADMMSCA}, for $M=50,100$, and $200$. We see that in all cases, the outer layer  converges  quickly in just a few iterations. Based on these convergence studies, for the rest of simulations, we set the inner-layer threshold to be $1 \times10^{-3}$ and that for the outer-layer SCA to be $1 \times10^{-3}$.

%%%%%%%%%%%%%%%%%%%%%%%%%%%%%%%%%%%%%%%%%%%%%%%%%%%%%%%%%
\subsection{Performance Comparison}\label{perQoS}
We now evaluate the performance of Algorithm~\ref{algADMMSCA}. For comparison, we also consider the following methods:
\begin{itemize}
\item 
OptSDR: Use the optimal beamforming structure  obtained in \eqref{eqn:thm1_w_opt}; Then, apply the conventional SDR method with Gaussian randomization to solve $\Pc_2$.
\item 
OptSCA-IPM: Use the optimal beamforming structure in \eqref{eqn:thm1_w_opt}; Then, apply the  SCA method to iteratively solve $\Pc_{2\SCA}(\ubf)$ via the standard convex solver CVX, which implement the interior-point method (IPM).
\item 
DirectSDR: Apply the SDR approach to $\Pc_{o}$ along with the Gaussian randomization method to compute $\{\wbf_i\}$ directly.
\item 
DirectSCA: Apply the SCA method to $\Pc_o$ by iteratively solving $\Pc_{1\SCA}(\Zbf)$ via the convex solver CVX to compute $\{\wbf_i\}$ directly.
\item 
Lower Bound for $\Pc_{o}$: Solve the relaxed problem of $\Pc_{o}$ via the SDR method directly. This is a benchmark for all the above methods.
\end{itemize}

 Note that we let OptSDR and OptSCA-IPM take the advantage of the optimal beamforming structure obtained in \eqref{eqn:thm1_w_opt} as well, but instead of our proposed  fast algorithm, we apply the conventional optimization techniques to compute  $\{\abf_i\}$, in order to compare the computational complexity of different optimization approaches. DirectSDR and DirectSCA are the conventional common methods in the literature to compute the beamforming vectors $\{\wbf_i\}$ directly, which require fully centralized processing. We consider these methods to evaluate the benefit of using the optimal structure.   

\begin{figure}[t]
\centering
\includegraphics[width=.9\columnwidth]{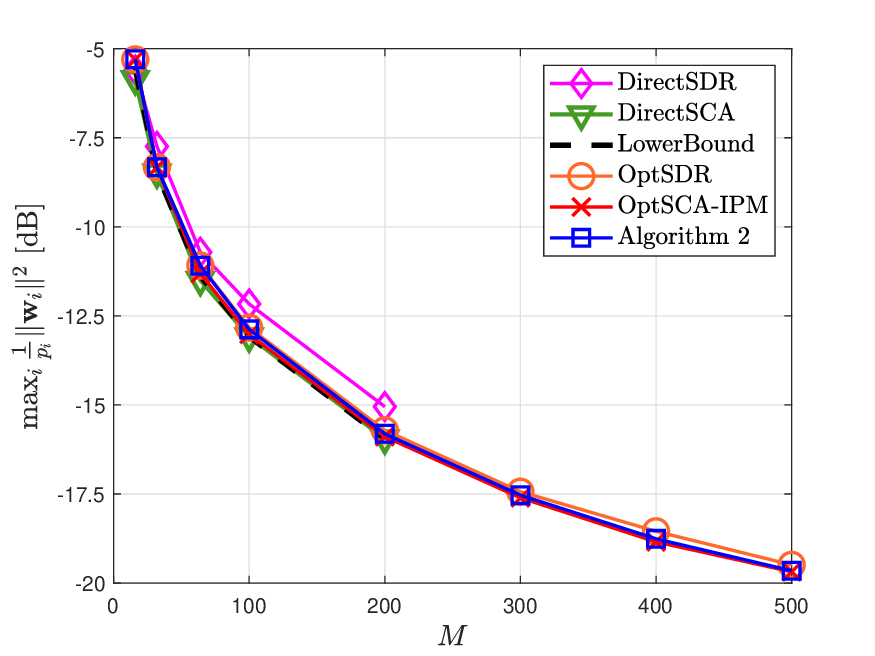}
\centering
\caption{ Average transmit power margin  $\max_i \frac{1}{p_i}\|\wbf_i\|^2$ vs. the number of antenna $M$ ($K = 5$, $J=3$).}\vspace*{-.5em}
\label{fig:perQoS}
\end{figure}

Fig.~\ref{fig:perQoS} shows the average maximum transmit power margin $\max_i \|\wbf_i\|^2/p_i$ vs. the number of antennas $M$. We set $K=5$ and  $\gamma_{ik} = 10$ dB, $\forall i,k$.
 Note that both DirectSDR and the lower bound incur very high computational complexity as $M$ becomes large, and their performance are only shown up to $M=200$. We see that the performance of Algorithm~\ref{algADMMSCA} is very close to the lower bound, suggesting that it  achieves a nearly-optimal performance. This indicates the effectiveness of our proposed approximate approach for computing $\lambdabf$ and the heuristic setting for $\mubf$ in Section~\ref{sec:Lambda&Mu}, and the computed solution based on the optimal beamforming structure is nearly optimal. The other methods also perform close to the lower bound, except for DirectSDR, which has a slight performance gap compared to the lower bound. 

Even though their performances are close,  the average computation times of these algorithms are substantially different, as shown in Table~\ref{tab:QoS_M}.\footnote{Note that in all our experiments, we did not use parallel computing for Algorithm~\ref{algADMMSCA} in computing   $\dbf_{ik}$'s and $\abf_i$'s. These quantities are computed sequentially instead. } The computation time of Algorithm~\ref{algADMMSCA}, OptSCA-IPM, and OptSDR remains roughly unchanged  as $M$ increases. This is because they  are all based on  the optimal beamforming structure in \eqref{eqn:thm1_w_opt} and only need to compute  weight vectors $\abf_i$'s with the total dimension $JK$, which is independent of $M$. This is in contrast to DirectSDR, whose  computation time increases with $M$ significantly as it computes $\wbf_i$'s directly, making it impractical for massive MIMO systems. Furthermore, the computational time of our proposed algorithm is several orders of magnitude lower than those of OptSCA-SDR and OptSCA-IPM. This  demonstrates the computational advantage of our proposed fast algorithm in Algorithm~\ref{algADMMSCA} based on the closed-form or semi-closed-form updates, as compared with the conventional convex solver. The communication overhead saving between BSs and the CPU by our semi-distributed computing approach in Algorithm~\ref{algADMMSCA} is shown in Table~\ref{tab:QoS_M_overhead}, where the amount of data exchange  by  Algorithm~\ref{algADMMSCA} is shown as a percentage of that of the conventional fully centralized processing using the full channel state information for different  $M$ values. We see that our approach  can substantially reduce the amount of data exchange over the fronthaul, especially when $M$ becomes large.  
\begin{table}[t]
  \renewcommand{\arraystretch}{1.2}
        \centering
        \caption{ Average computation time (sec.) ($J=3,K=5$).}\vspace*{-1em}
        \label{tab:QoS_M}
        \begin{tabular} {r|c|c|c|c|c}%adjust the space for columns
                \hline
                $M$& 100& 200& 300& 400& 500\\
                \hline\hline
%                $\lambdabf$& 0.0889&0.0573&0.1496&0.2433&0.3648\\
%                \hline
                Algorithm~\ref{algADMMSCA}& 0.039             &0.032&0.058&0.047&0.043\\
                \hline
                OptSCA-IPM& 3.74  & 3.64& 3.48& 3.42&3.89\\
                \hline
                OptSDR& 0.61&0.61&0.65&0.60&0.70\\
                \hline
                DirectSDR& 33.9&239&--&--&--\\
                \hline
        \end{tabular}\vspace*{1em}        
%\end{table}
%\begin{table}[t]
  \renewcommand{\arraystretch}{1.2}
        \centering
        \caption{ Communication Overhead of Proposed over Fully Centralized   ($J=3,K=5$).}\vspace*{-1em}
        \label{tab:QoS_M_overhead}
        \begin{tabular} {p{.8in}|c|c|c|c|c}%adjust the space for columns
                \hline
                \hspace*{4em}$M$& 100& 200& 300& 400& 500\\
                \hline\hline
               Semi-distributed (Algorithm~\ref{algADMMSCA})   & 9.7\%&4.8\%&2.9\%&2.2\%&1.7\%\\
                \hline
        \end{tabular}\vspace*{-1em}
\end{table}

\begin{figure}[t]
\centering
\includegraphics[width=.9\columnwidth]{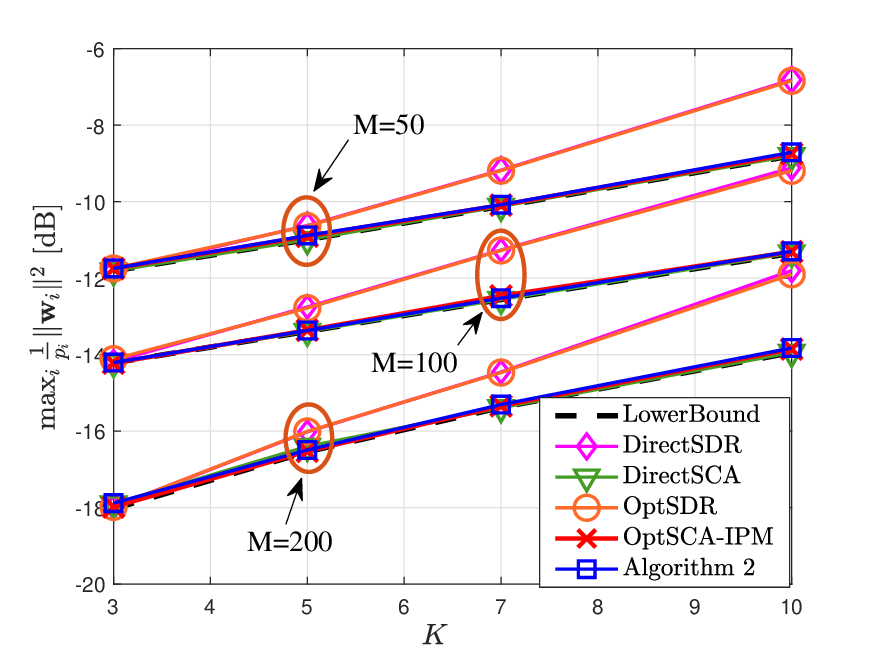}
\centering
\caption{ Average transmit power margin  $\max_i \frac{1}{p_i}\|\wbf_i\|^2$ vs. 
$K$ users per cell in the two-tier 19-cell setup. ($M = 100$, $J=3$).} \vspace*{-.5em}
\label{fig:perQoSK}
\end{figure}

To study the effect of the number of users on the performance, in Fig.~\ref{fig:perQoSK}, we show $\max_i \|\wbf_i\|^2/p_i$  vs.  $K$ users per group for $M=50,100,200$. We see that Algorithm~\ref{algADMMSCA} and OptSCA-IPM can nearly attain the lower bound for all values of $K$ and $M$. However, OptSDR deteriorates substantially as $K$ increases, with a noticeable $\sim 2$ dB gap to the lower bound for $K=10$. This is expected for the SDR-based method, which is an approximation method  known to be less accurate as the  problem size increases, particularly the number of constraints.

Table~\ref{tab:QoS_K} shows the average computation time of these methods as $K$ increases.  We see that the computation time of Algorithm~\ref{algADMMSCA}
increases only mildly as $K$ grows and is several orders of
magnitude lower than other methods. This again demonstrates
the computational advantage of Algorithm~\ref{algADMMSCA} over other
methods. Its scalability is highly desirable for massive MIMO systems.
 Table~\ref{tab:QoS_K_overhead} shows the amount of data exchange  by  Algorithm~\ref{algADMMSCA}  as a percentage of that of the conventional fully centralized processing for different  $K$ values. We again see that the required data exchange in our approach is only a small fraction of that needed for fully centralized processing. 

\begin{figure}[t]
\centering 
\includegraphics[width=.9\columnwidth]{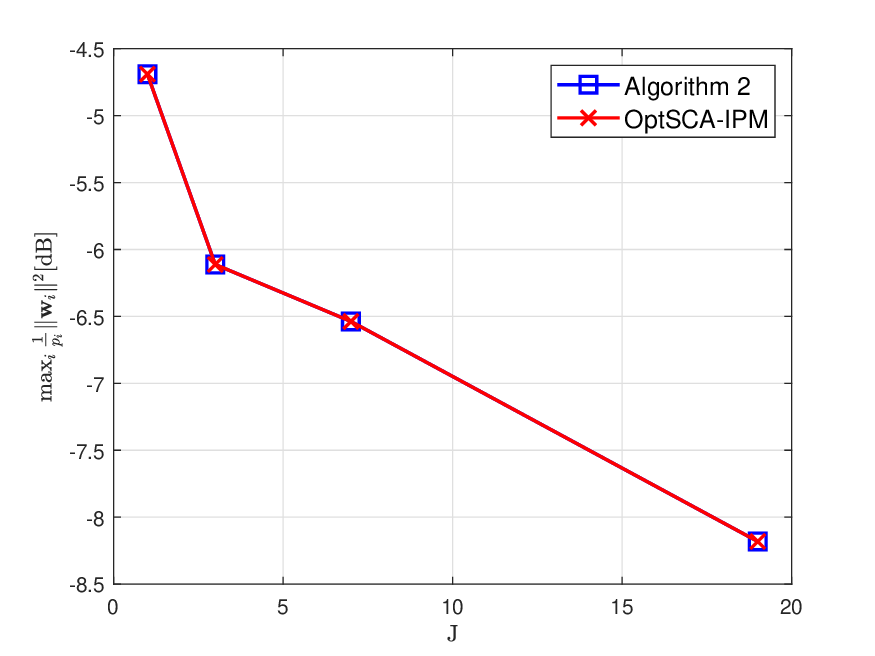}
\centering
\caption{Average transmit power margin  $\max_i \frac{1}{p_i}\|\wbf_i\|^2$ vs. the number of coordinating cells $J$ ($K = 5$, $M=100$).}\vspace*{-.5em}
\label{fig:perJ}
\end{figure}

\begin{table}[t]
        \renewcommand{\arraystretch}{1.2}
        \centering
        \caption{ Average computation time (sec.) ($M=100, J=3$).}\vspace*{-1em}
        \label{tab:QoS_K}
        \begin{tabular} {r|c|c|c|c}%adjust the space for columns
                \hline
                $K$& 3& 5& 7& 10\\
                \hline\hline
                Algorithm~\ref{algADMMSCA}& 0.0057&0.041&0.17  &0.44\\
                \hline
                OptSCA-IPM& 1.24&3.80&4.71&9.52\\
                \hline
                OptSDR& 0.52&0.56&0.72&1.02
                \\
                \hline
                DirectSDR& 45.8&217&375&1056\\
                \hline 
        \end{tabular}\vspace*{1em}
          \renewcommand{\arraystretch}{1.2}
        \centering
        \caption{Communication Overhead of Proposed over Fully Centralized   ($M=100, J=3$).}\vspace*{-1em}
        \label{tab:QoS_K_overhead}
        \begin{tabular} {p{.75in}|c|c|c|c}%adjust the space for columns
                \hline
                \hspace*{4em}$K$& 3& 5& 7& 10\\
                \hline\hline
               Semi-distributed (Algorithm~\ref{algADMMSCA}) & 7\%&9.7\%&12.3\%&16.3\%\\
                \hline
 \end{tabular}\vspace*{-1em}
\end{table}
Finally, we examine the effect of varying the number of coordinating BSs $J$. We consider a two-tier cell setup consisting of $19$ cells and vary the number of coordinating cells as  $J=1, 3, 7, 19$.  We  consider a practical cellular network configuration where the cell radius is $500$ m. The channel path loss is modeled as $139.1+35\log10(d_{ijk})$, where $d_{ijk}$ is the distance of BS $i$ to user $k$ in cell $j$ in km.    The system bandwidth is 10 MHz, and the receiver noise is $-94$ dBm. We set the power budget for each BS to $p_i = 45$ dBm, and SINR target $\gamma_{ik}=15$~dB.  The  performances of  Algorithm~\ref{algADMMSCA} and OptSCA-IPM are shown in Fig.~\ref{fig:perJ}, for $M=100$ and $K=5$. We see that the two algorithms  perform nearly identically. The BS transmit power decreases as  $J$ increases, due to improved interference management with more coordinating BSs.  The computation times for both algorithms are shown in Table~\ref{tab:QoS_J}. Note that the parallel computing approach for Algorithm~\ref{algADMMSCA}, which is discussed in Section~\ref{subsec:complexity}, was not utilized in these simulation results; instead, all updates were  computed sequentially. Despite this, Table V demonstrates that the average computation time of Algorithm~\ref{algADMMSCA} remains  low as $J$ increases to  $19$ cells. In contrast, the computation time for OptSCA-IPM significantly increases when $J$ reaches $19$. 

\begin{table}[t]
        \renewcommand{\arraystretch}{1.2}
        \centering
        \caption{Average computation time (sec.) ($M=100$, $K=5$).}\vspace*{-1em}
        \label{tab:QoS_J}
        \begin{tabular} {r|c|c|c|c}%adjust the space for columns
                \hline
                $J$& 1& 3& 7& 19\\
                \hline\hline
                Algorithm~\ref{algADMMSCA}& 0.0019&0.039&0.30  &2.78\\
                \hline
                OptSCA-IPM& 0.32&3.4&9.0&122.3\\
                \hline 
        \end{tabular}\vspace*{-1em}
\end{table}

%%%%%%%%%%%%%%%%%%%%%%%%%%%%%%%%%
\section{Conclusion}\label{sec:conclusion}
This paper considers BSs coordination for   multicast beamforming   and provides a computation-communication efficient   solution for massive MIMO cellular networks. Considering the QoS problem for individual BS transmit power minimization, we first obtain the optimal coordinated multicast beamforming structure. It shows that the optimal beamformers are naturally distributed beamformers, each being a  function of
the local CSI at its BS only.
Furthermore, the beamforming solution has an inherent low-dimensional structure, where the essential unknown weights  to be determined are in the dimension of the serving users at each BS, which are independent of the number of BS antennas. We judiciously explore this optimal structure and propose a scalable and fast algorithm with a semi-distributed computing approach for BSs to determine their beamformers based on the local CSI and limited essential information
sharing, thus significantly reducing the fronthaul communication load for coordination in massive MIMO networks.
We further show that the beamforming structural results and our algorithm can be extended to imperfect CSI case and the scenario involves BS clustering for full cooperation. Simulation results show that our proposed algorithm built upon the optimal structure 
achieves a near-optimal performance and is scalable to the network size with substantially lower computational complexity and communication overhead than other alternatives.

% For the future work, the optimal structure and the first-order techniques can be extended into multi-cell multi-group multicast massive MIMO systems. Compared with our system model, there are more than one groups located in each cell, which make the model become more complex. Instead of only considering the maximum power among BSs, the objective also includes the total power consumed by each group. These extra constraints may cause the optimal structure not be exactly same as ours, but the derivation process and technique should still be able to utilize to obtain the optimal solution.
    
\appendices
\section{Proof of Proposition~\ref{prop1}}\label{app_prop1}
\IEEEproof
Under the optimal Lagrange multipliers $(\lambdabf^{\star},\mubf^{\star})$ for the dual problem $\Dc_{1\SCA}(\Zbf)$,  we have the following KKT condition for the minimization of $\Lc(\Wbf,t,\lambdabf^{\star},\mubf^{\star};\Zbf)$ in \eqref{eqn:QoSproblemgFunc}, 
\begin{align}
\frac{\partial \Lc(\Wbf,t,\lambdabf^{\star},\mubf^{\star};\Zbf)}{\partial \wbf_i^H}  &= \Rbf_{i^{-}}(\lambdabf^{\star}\!,\mubf^{\star})\wbf_{i}(\Zbf)-\nubf_{i} \!= 0. \label{eqn:Prop1_w} \\
\frac{\partial \Lc(\Wbf,t,\lambdabf^{\star},\mubf^{\star};\Zbf)}{\partial t} &=1- \mathbf{1}^T\mubf^\star=0. \label{eqn:Prop1_mu}
\end{align} 
 From \eqref{eqn:Prop1_mu}, $\mathbf{1}^T\mubf^\star=1$. For \eqref{eqn:Prop1_w}, we now show that $\Rbf_{i^{-}}(\lambdabf^{\star},\mubf^{\star})$ in \eqref{eqn:QoSOptStrR-} is invertible.  

We  discuss this in two cases. \emph{i) $M>(J-1)K$:} We first consider the typical case of system setup where the number of BS antennas is more than the number of users in other coordinating cells\footnote{In the typical system operation, there are more BS antennas than the available active users for interference management.}. We employ proof by contradiction. Assume $\mu_i^\star=0$ for some $i$. Then, $\Rbf_{i^{-}}(\lambdabf^{\star},\mubf^{\star})$ in \eqref{eqn:QoSOptStrR-} is rank deficient. Notice that the range of   $\Rbf_{i^{-}}(\lambdabf^{\star},\mubf^{\star})$ is spanned  by channels from BS $i$ to all users in other cells, $\{\hbf_{i,jk}, k\in \Kc, j \in \Jc, j \neq i\}$, while  $\nubf_i$ from \eqref{eqn:QoSOptStrNu} is a linear combination of channels from BS $i$ to its own users in cell $i$: $\{\hbf_{i,ik},k\in\Kc\}$. Note that all user channels are random realizations following certain channel distributions, and the channels of out-of-cell users are independent of in-cell users. Thus,  with probability 1 (w.p.1.) that  $\nubf_i$ does not lie in the range of $\Rbf_{i^{-}}(\lambdabf^{\star},\mubf^{\star})$.  Then,   there is no solution to the linear equation in \eqref{eqn:Prop1_w}  for $\wbf_i(\zbf)$. This means that the partial derivative in \eqref{eqn:Prop1_w} will not be 0 at optimality. This contradict with the KKT condition of the optimal  solution to $\Pc_{1\SCA} (\zbf)$. Thus, the optimal $\mu_i^\star>0$, $i\in \Jc$, and $\Rbf_{i^{-}}(\lambdabf^{\star},\mubf^{\star})$ is invertible. \emph{ii) $M \leqslant(J-1)K$:} In this less likely scenario with insufficient number of antennas available, as mentioned earlier, all user channels are random channel realizations, and thus, the second term of $\Rbf_{i^{-}}(\lambdabf^{\star},\mubf^{\star})$ in \eqref{eqn:QoSOptStrR-} has a full rank (w.p.1), and we can directly conclude that $\Rbf_{i^{-}}(\lambdabf^{\star},\mubf^{\star})$  is  invertible in this case. 
  
Following the above, from \eqref{eqn:Prop1_w} we have 
\vspace*{-.5em}
\begin{align*}
\wbf^{\star}_{i}(\Zbf)&=\Rbf^{-1}_{i,i^{-}}(\lambdabf^{\star},\mubf^{\star})\bigg(\sum^{K}_{k=1}
\lambda_{ik}^\star\hbf_{i,ik}\hbf^{H}_{i,ik}\bigg)\zbf_{i} \nn\\
&=\Rbf^{-1}_{i,i^{-}}(\lambdabf^{\star},\mubf^{\star}) \sum^{K}_{k=1}\left(\lambda_{ik}^\star\hbf^{H}_{i,ik}\zbf\right)\hbf_{i,ik},
\end{align*}
which leads to \eqref{eqn:OptStruc1QoS}. \endIEEEproof

%%%%%%%%%%%%%%%%%%%%%%%%%%%%%%%%%%%%%%%%%%%%%%%
\section{Proof of Theorem~\ref{theorem-1QoS}}\label{appA}
\IEEEproof
The proof follows the  technique used in the proof of \cite[Theorem \ref{theorem-1QoS}]{Dong&Wang:SP20}. Specifically, following the optimal $\wbf^{\star}_{i}(\Zbf)$ for $\Pc_{1\SCA}(\Zbf)$ in \eqref{eqn:OptStruc1QoS}, we have 
\vspace*{-.5em} 
\begin{align}
\label{eqn:relationRi-wi}
\Rbf_{i,i^{-}}(\lambdabf^{\star},\mubf^{\star})\wbf^{\star}_{i}(\Zbf) = \sum_{k=1}^K
\lambda_{ik}^\star\hbf_{i,ik}\hbf_{i,ik}^H \zbf_i.
\end{align}
From $\Rbf_{i}(\lambdabf^{\star},\mubf^{\star})$ in \eqref{eqn:covarianceMatrixRwithMuQoS}, we have 
\begin{align}
\label{eqn:RiTimesW}
&\Rbf_{i}(\lambdabf^{\star},\mubf^{\star})\wbf^{\star}_{i}(\Zbf)\nn\\
&= \bigg( \Rbf_{i,i^{-}}(\lambdabf^{\star},\mubf^{\star})+ \sum_{k = 1}^{K} 
\lambda_{ik}\gamma_{ik}\hbf_{i,ik}\hbf_{i,ik}^H  \bigg)\wbf^{\star}_{i}(\zbf) \nn \\ 
&\stackrel{(a)}{=} \sum_{k=1}^{K} \lambda_{ik}^{\star}\hbf_{i,ik}\hbf_{i,ik}^H \zbf_{i} + 
\sum_{k=1}^{K} \lambda_{ik}^{\star}\gamma_{ik}\hbf_{i,ik}\hbf_{i,ik}^H 
\wbf^{\star}_{i}(\zbf) \nn \\
&= \sum_{k = 1}^{K}\lambda_{ik}^{\star}(1+\gamma_{ik})\left(\hbf_{i,ik}\hbf_{i,ik}^{H} \zbf_{i}+\hbf_{i,ik}\hbf_{i,ik}^{H} \wbf^{\star}_{i}(\zbf)\right)
\end{align}
where $(a)$ follows the equation in \eqref{eqn:relationRi-wi}.
Assume the initial $\Zbf^{(0)}$  in the SCA procedure is close to the  global optimal solution, and the  SCA iteration  converges to the global optimal solution, \ie  $\Zbf \rightarrow \Wbf^{o}$. Then, we have $\wbf^{\star}_{i}(\Zbf) \rightarrow \wbf^{o}_{i}$. Following this, we have $\hbf_{i,ik}^{H} \zbf_i \to \hbf_{i,ik}^{H} \wbf^o_i$, and $\hbf_{i,ik}^{H} \wbf^{\star}_{i}(\Zbf) \to \hbf_{i,ik}^{H} \wbf^o_i$.
Also, as $\wbf^{\star}_{i}(\Zbf) \rightarrow \wbf^o_i$, the optimal $(\lambdabf^{\star},\mubf^\star)$ for the dual problem $\Dc_{1\SCA}(\Zbf)$ also converges to the optimal $(\lambdabf^o,\mubf^o)$ of $\Dc_{1\SCA}(\Wbf^o)$, which is the dual problem of $\Pc_1$. Thus, at the limit of $\zbf_{i} \rightarrow \wbf^{o}_i$,   \eqref{eqn:RiTimesW} becomes
\begin{align*}
\Rbf_{i}(\lambdabf^{o}, \mubf^{o})\wbf^{o}_{i} & = 
\sum_{k = 1}^{K}\lambda_{ik}^{o} (1+\gamma_{ik})(\hbf_{i,ik}^{H} \wbf^o_i) \hbf_{i,ik} =\Hbf_{i}\abf^{o}_{i} 
\end{align*}
where $a_{ik}^o=\lambda_{ik}^{o} (1+\gamma_{ik})(\hbf_{i,ik}^{H} \wbf^o_i)$. Following the  argument in Appendix~\ref{app_prop1}, we can similarly show that $\Rbf_{i}(\lambdabf^{o}, \mubf^{o})$ is full rank and invertible. Thus, we obtain the optimal solution  $\wbf_i^o$ in \eqref{eqn:thm1_w_opt}. 

The optimal objective value of $\Pc_o$ is  the optimal $t^o$ in $\Pc_1$. In each SCA iteration, the optimal solution   $\wbf^{\star}_{i}(\Zbf) $ to $\Pc_{1\SCA}(\Zbf)$ is given in  \eqref{eqn:OptStruc1QoS}. We can rewrite  it in a compact form as follows:
\begin{align}
\label{eqn:OptStruc2QoSMatrix}
\wbf^{\star}_{i}(\Zbf)= \Rbf^{-1}_{i,i^{-}}(\lambdabf^{\star},\mubf^{\star})
\Hbf_{i}\Dbf_{\lambdabf_{i}}\Hbf^{H}_{i}\zbf_{i}.
\end{align}
where $\Dbf_{\lambdabf_{i}}\triangleq \text{diag}(\lambdabf_{i})$. Substituting the expression of    $\wbf^{\star}_{i}(\Zbf) $  in   \eqref{eqn:OptStruc2QoSMatrix} into  \eqref{eqn:QoSLagrangianRegrouping}, the dual function in \eqref{eqn:QoSproblemgFunc} can be written as 
\begin{align} \label{eqn:QoSgFuncDerive}
g(& \lambdabf,\mubf;\Zbf) \nn \\
&=  \Big(\!1\!-\!\sum^{J}_{i=1}\mu_{i}\Big)t^\star+\sigma^{2}\sum^{J}_{i=1}\lambdabf^{T}_{i}\gammabf_{i}+
\sum^{J}_{i=1}\zbf^{H}_{i}\Hbf_{i}\Dbf_{\lambdabf_{i}}\Hbf^{H}_{i}\zbf_{i} \nn \\
&\quad -\sum^{J}_{i=1}\zbf^{H}_{i}\Hbf_{i}\Dbf_{\lambdabf_{i}}\Hbf^{H}_{i}\Rbf^{-1}_{i,i^{-}}
(\lambdabf^{\star},\mubf^{\star})\Hbf_{i}\Dbf_{\lambdabf_{i}}\Hbf^{H}_{i}\zbf_{i} \nn \\
&=\sum^{J}_{i=1}\zbf^{H}_{i}\Hbf_{i}\Dbf_{\lambdabf_{i}}\Hbf^{H}_{i}
\left(\Ibf-\Rbf^{-1}_{i,i^{-}}(\lambdabf^{\star},\mubf^{\star})\Hbf_{i}
\Dbf_{\lambdabf_{i}}\Hbf^{H}_{i}\right)\zbf_{i} \nn \\
&\quad +   \Big(\!1\!-\!\sum^{J}_{i=1}\mu_{i}\Big)t^\star +\sigma^{2}\sum^{J}_{i=1} \lambdabf^{T}_{i}\gammabf_{i}.
\end{align}
where $\gammabf_i$ is defined below \eqref{QoS:opt_value}.
 Since the optimal solution $\wbf^{o}_i$ is a stationary solution, it can also be rewritten as   \eqref{eqn:OptStruc1QoS}   
\begin{align*}
\wbf^{o}_{i}&=\Rbf^{-1}_{i^-}(\lambdabf^{o}, \mubf^{o})\Hbf_{i}\alphabf^{o}_{i}.
\end{align*}

If the SCA iteration converges the optimum $\zbf_i \rightarrow \wbf^o_i$, we have $\hbf_{i,ik}^{H} \zbf_i \to \hbf_{i,ik}^{H} \wbf^o_i$, $\lambda_{ik}^\star\to \lambda_{ik}^o$, and  $\alpha_{ik}^\star \to \alpha_{ik}^o=\lambda_{ik}^o\hbf_{i,ik}^{H} \wbf^o$. Thus, $\alphabf_i^o=\Dbf_{\lambdabf_i^{o}} \Hbf^{H}_{i}\wbf^{o}$. Substituting  it into the above expression, we have 
$\wbf^{o}_{i}=\Rbf^{-1}_{i^-}(\lambdabf^{o}, \mubf^{o})\Hbf_{i} \Dbf_{\lambdabf^{o}} \Hbf^{H}_{i}\wbf^{o}_{i}$, which leads to 
$\left(\Ibf-\Rbf^{-1}_{i,i^{-}}(\lambdabf^{o},\mubf^{o})\Hbf_{i}
\Dbf_{\lambdabf^{o}_{i}}\Hbf^{H}_{i}\right)\wbf^{o}_{i} = \bf{0}$.
Following this equation and since $\zbf_{i}\rightarrow \wbf^{o}_{i}$,  the first term in \eqref{eqn:QoSgFuncDerive}  will be $0$ at optimality. Also, since  $\mathbf{1}^{T}\mubf^{\star}= 1$  in    \eqref{eqn:OptStruc1QoS}. We have $\mathbf{1}^{T}\mubf^{o}= 1$  as well. Thus, the second term in \eqref{eqn:QoSgFuncDerive} will be $0$ at optimality. It follows that as  $\zbf\rightarrow \wbf^{o}$, we have
\vspace*{-.5em}
\begin{align}
\label{eqn:QoSLagrangianOptValue}
\max_{\lambdabf,\mubf}g( \lambdabf,\mubf;\Wbf^o) = \sigma^{2}\sum^{J}_{i=1} \lambdabf^{oT}_{i}\gammabf_{i} = \sigma^{2}\lambdabf^{o T}\gammabf.
\end{align}
Also, we have $\Pc_{1\SCA}(\Zbf)\rightarrow \Pc_1$ ($\Pc_o$). Thus, the minimum  objective value of $\Pc_o$ is given by \eqref{eqn:QoSLagrangianOptValue}. 
\endIEEEproof

%%%%%%%%%%%%%%%%%%%%%%%%%%%%%%%%%%%%%%%%%%%%
\section{The Solution to $\Pc_{\dbf\text{sub}}(\ubf)$ in \eqref{eqn:ADMMdSubQoS}}\label{appD}
Define $e^{(l)}_{1, j,ik} \triangleq \abf_{j}^{(l)H} \fbf_{j,ik} - q^{(l)}_{j,ik}$, $e_{2, ik} \triangleq |\ubf^{H}_{i} \fbf_{i,ik}|^{2}+\gamma_{ik}\sigma^{2}$, $e_{3, ik} \triangleq \ubf_{i}^H\fbf_{i,ik}$. Then, 
%\begin{align}
% \label{eqn:ADMMdSubQoS}
% &\Pc_{\textrm{dsub}}(\ubf): \  \min_{\dbf_{ik}} \frac{\rho}{2}\sum_{j=1}^{J}
% |d_{j,ik} - e^{(l)}_{1,j,ik}|^{2} \nn \\
% & \quad \st \ e_{2, ik} + \gamma_{ik}\sum^{J}_{j=1,j\neq i} |d_{j,ik}|^{2} - 2\mathfrak{Re}\{ d_{i,ik} e_{3, ik}\} \leqslant 0
% \end{align}
the optimal solution $\dbf_{ik}^{o}$ for $\Pc_{\dbf\text{sub}}(\ubf)$ is given by 
\begin{align}
d_{j,ik}^{o} = \begin{cases}e_{1, i,ik}^{(l)}+\nu_{ik}^{o}e_{3, ik}, & j=i, \\
\frac{e^{(l)}_{1,j,ik}}{1+\nu_{ik}^{o}\gamma_{ik}}, & j \neq i \\
\end{cases}
\label{eqn:ADMM-dj,ikQoS}
\end{align}
where $\nu_{ik}^{o}\geqslant 0$ is the optimal Lagrange multiplier associated with the constraint in \eqref{eqn:ADMMdSubQoS}.
Substituting  $d_{j,ik}^{o}$ in \eqref{eqn:ADMM-dj,ikQoS} into the  constraint in \eqref{eqn:ADMMdSubQoS} leads to 
\begin{align}
\label{eqn:inequalADMMderiveQoS}
f(\nu_{ik}^{o}) \triangleq& \ e_{2,ik}+\gamma_{ik}\frac{\sum^{J}_{j \neq i}|e_{1,j,ik}^{(l)} |^{2}}{(1+\nu_{ik}^{o}\gamma_{ik})^{2 }}-2\mathfrak{Re} \{e^{(l)}_{1,i,ik}e_{3,ik}^* \}\nn \\
&-2\nu^{o}_{ik}| e_{3,ik} |^{2} \leqslant 0, 
\end{align}
which is strictly decreasing for $\nu_{ik}^o \geqslant 0$. The solution  $\nu_{ik}^o$ is obtained as $\nu_{ik}^o = 0$ if $e_{2,ik}+\gamma_{ik}\sum^{J}_{j \neq i}|e_{1,j,ik}^{(l)} |^{2}-2\mathfrak{Re} \{e^{(l)}_{1,i,ik}e_{3,ik}^* \}\leqslant 0$; otherwise, it is the unique positive root of $f(\nu_{ik}^o)=0$, which has a closed-form cubic formula.

\bibliographystyle{IEEEtran}
\bibliography{Yin,myBib}
\end{document}